\documentclass[aps,pra,reprint,10pt,a4paper]{revtex4-1}
\usepackage[utf8]{inputenc}
\usepackage[sc,osf]{mathpazo}
\usepackage{amsmath}
\usepackage[T1]{fontenc}
\usepackage{latexsym}
\usepackage{amssymb}
\usepackage[colorlinks=true,citecolor=blue,urlcolor=blue]{hyperref}
\usepackage{color}
\usepackage{graphics,epstopdf}
\usepackage{soul}
\usepackage[demo]{graphicx}
\usepackage{capt-of}
\usepackage{lipsum}
\usepackage{adjustbox}
\usepackage[normalem]{ulem}
\usepackage[table,xcdraw]{xcolor}
\usepackage{braket}
\usepackage{physics}
\usepackage{ragged2e}

\usepackage{comment}
\usepackage[font=small,labelfont=bf,justification=justified,singlelinecheck=false]{caption}

\begin{document}

\title{Qubit magic-breaking channels\\
}

\author{ Ayan Patra$^1$, Rivu Gupta$^{1,2}$, Alessandro Ferraro$^{2,3}$, Aditi Sen(De)$^1$}

\affiliation{$^1$ Harish-Chandra Research Institute,  A CI of Homi Bhabha National Institute, Chhatnag Road, Jhunsi, Prayagraj - $211019$, India\\
$^2$ Dipartimento di Fisica “Aldo Pontremoli,” Università degli Studi di Milano, I-$20133$ Milano, Italy\\
$^3$ Centre for Quantum Materials and Technologies, School of Mathematics and Physics,
Queen’s University Belfast, BT$7$ $1$NN Belfast, United Kingdom}

\begin{abstract}

We develop a notion of quantum channels that can make states useless for universal quantum computation by destroying their magic (non-stabilizerness) --- we refer to them as magic-breaking channels. We establish the properties of these channels in arbitrary dimensions. We prove the necessary and sufficient criteria for qubit channels to be magic-breaking and present an algorithm for determining the same. Moreover, we provide compact criteria in terms of the parameters for several classes of qubit channels to be magic-breaking under various post-processing operations. 
Further, we investigate the necessary and sufficient conditions for the tensor product of multiple qubit channels to be magic-breaking. We establish implications of the same for the dynamical resource theory of magic preservability.

\end{abstract}

\maketitle
\section{Introduction}
\label{sec:intro}

Quantum channels~\cite{nielsen_2010, Preskill, wilde_2013, Watrous_2018}, defined as completely positive and trace-preserving (CPTP) linear maps~\cite{Ruskai_LAA_2002, krauss_2013} which take density matrices from the input Hilbert space to those in the output Hilbert space~\cite{dePillis_PJM_1967}, represent the medium transmitting quantum information during the implementation of information-theoretic protocols. Hence, the intrinsic properties of a quantum channel can impact any protocol it comprises, thereby making their characterization of fundamental importance. Notable works in this direction include the isomorphism between states and channels~\cite{Jamiolkowski_RMP_1972, Choi_LAA_1975, Paulsen_JMP_2013, Kye_JMP_2022}, information carrying capacity~\cite{Holevo_PPI_1973, Holevo_PPI_1973_2, Holevo_IEEE_1998, Schumacher_PRA_1997, LLoyd_PRA_1997}, superadditivity~\cite{Bennett_PRL_1997, DiVincenzo_PRA_1998, Hastings_Nature_2009}, and superactivation~\cite{Smith_Science_2008, Cubitt_IEEE_2011, Brandao_IEEE_2013, Shirokov_CMP_2015} of information in both finite~\cite{Bennett_IEEE_2002, Devetak_IEEE_2005, Shirokov_TVP_2008, Shirokov_PRA_2018, Shirokov_JMP_2019} and infinite dimensions~\cite{Holevo_TPA_2006, Holevo_PIT_2008, Shirokov_PIT_2008, Holevo_DM_2010}.
Moreover, various kinds of quantum channels have been identified, such as unital channels~\cite{Landau_LAA_1993, Mendl_CMP_2009, Haagerup_AHP_2021, Girard_CMP_2022, Li_arXiv_2023}, which play a vital role in state discrimination~\cite{Bae_JPA_2015} and operator error correction~\cite{Kribs_PRL_2005, Poulin_PRL_2005, Bacon_PRA_2006, Nielsen_PRA_2007}, and group-covariant channels~\cite{Memarzadeh_PRR_2022, Mozrzymas_JMP_2017, Kopszak_JPA_2020}, used to establish the strong converse theorem of channel coding~\cite{Konig_PRL_2009, Datta_QIP_2016} and in studying the relative entropy bound of private communication~\cite{Wilde_IEEE_2017}.

A novel way to characterize quantum channels is based on their ability to destroy the quantum resources~\cite{Liu_PRL_2017} inherent in quantum states. Notable examples of such resource-eradicating channels include entanglement-~\cite{Ruskai_RMP_2003, Horodecki_RMP_2003, Ziman_JPA_2010, Filipov_PRA_2012, Filipov_PRA_2013}, non-locality-~\cite{Pal_JPA_2015}, and coherence-breaking channels~\cite{Luo_QIP_2022}, which produce outputs devoid of the respective quantum resource. Instead of resource-breaking channels, the notion of ones that render quantum states, initially useful for information-processing tasks, useless after being sent through them, thereby prohibiting quantum advantage, has been recently proposed and referred to as process-resource-breaking trace-preserving channels (PBT)~\cite{Muhuri_arXiv_2023}. A rigorous analysis of this framework has been conducted based on the communication protocols of dense coding~\cite{Bennett_PRL_1992} and teleportation~\cite{Bennett_PRL_1993}.

Beyond quantum information-theoretic protocols, another thriving area is the theory of universal quantum computation~\cite{Reck_PRL_1994, Deutch_PRSL_1995, DiVincenzo_PRA_1995, Lloyd_PRL_1995, Boykin_IEEECS_1999}, and to identify resources that can leverage exponential speedup as compared to the best available classical algorithms~\cite{nielsen_2010}. Since quantum states are fragile under environmental influences, it is essential to develop a fault-tolerant scheme for implementing quantum algorithms~\cite{Grover_ACM_1996, Shor_SIAM_1997, Jozsa_PRSL_1998} such that they can provide an advantage even in the presence of decoherence \cite{Shor_IEEE_1996}. The stabilizer theory~\cite{Gottesman_PRA_1996} of fault-tolerant quantum computation~\cite{DiVincenzo_PRL_1996, Gottessman_PRA_1998} aims to address this question by defining quantum operations that do not spread errors and are hence transversal. However, it has been established that fault-tolerant schemes that are both universal and transversal do not exist~\cite{Eastin_PRL_2009}. In fact, the Gottesman-Knill theorem shows that stabilizer codes are not only sub-universal but also classically simulable~\cite{Gottesman_1999}. The leading approach to circumvent this problem and unlock universality is to use non-stabilizer states and operations, also known as magic states~\cite{Shor_IEEE_1996, Gottesman_Nature_1999}. Therefore, non-stabilizerness is necessary to obtain quantum advantage through universal quantum computation~\cite{mari2012positive, veitch2012negative}, leading to the development of a specialized resource theory of magic~\cite{Veitch_NJP_2014}.

In this work, we aim to identify channels that can destroy the inherent resource of magic from states essential for universal computation. In particular, we introduce \textit{magic-breaking channels}, that yield resourceless stabilizer states as output from input magic states. In this sense, these channels belong to the set of PBT channels, where the process considered is universal quantum computation (c.f. Ref~\cite{Muhuri_arXiv_2023} for quantum communication). We prove the properties of magic-breaking channels independent of the dimension of the Hilbert space and identify a class of extreme points. Moreover, we obtain the necessary and sufficient conditions for qubit channels to be magic-breaking. Further, we develop an algorithm to determine such channels and elucidate different classes of qubit magic-breaking channels characterized by different post-processing operations. We also establish a connection between qubit entanglement-breaking and magic-breaking channels even though they destroy resources of fundamentally different characters. We also prove the necessary conditions for multi-qubit magic-breaking channels and investigate their sufficiency. We discuss their implications for analyzing a dynamical resource theory of magic preservability.

The paper is organized in the following way.  In Sec.~\ref{sec:magic-breaking} we first enumerate properties of magic-breaking channels in arbitrary finite dimensions. Sec.~\ref{sec:qubit_magic-breaking} contains our main results where we rigorously characterize different classes of qubit-magic-breaking channels by following an algorithm for their identification. Finally, by considering multi-qubit states in Sec.~\ref{sec:multi-qubit_MB}, we lay down conditions under which a tensor product of single-qubit magic-breaking channels can destroy the resource in multi-qubit magic states. We conclude our paper with discussions in Sec.~\ref{sec:conclu}.

\section{Magic-breaking channels and their properties}
\label{sec:magic-breaking}

Here, we introduce the concept of magic-breaking channels, i.e., the channels that transform magic states into stabilizer states. Let us first define the set of stabilizer states which are considered to be non-resourceful for universal quantum computation.

\subsubsection{The stabilizer states}

The pure stabilizer states are defined by the action of the Clifford unitaries, which map the Weyl-Heisenberg operators to themselves (up to a phase), on the zero computational basis state, i.e., $\{S_i\} = \{U_i\ket{0}\}~\forall~U_i \in \text{Clifford}$ (see Appendix. \ref{subsubsec:stab} for a preliminary discussion on Weyl-Heisenberg operators). The stabilizer states, $\Xi$, on a $d$-dimensional Hilbert space, $\mathbb{C}^d$, belong to the convex hull of $\{S_i\}$,

\begin{eqnarray}
    \text{STAB}(\mathbb{C}^d) = \{\Xi : \Xi = \sum_i p_i S_i\},
    \label{eq:stabilizer_set}
\end{eqnarray}
with $p_i$ representing probabilities. The stabilizer states form a convex polytope~\cite{Veitch_NJP_2014} in the state space. Stabilizer operations, crucial for fault-tolerant quantum computing, comprise the preparation of stabilizer states, computational basis measurement, and Clifford rotations~\cite{Veitch_NJP_2014}. Such operations are efficiently simulable on a classical computer~\cite{Gottesman_1999} and hence cannot be used for universal quantum computation. The universality is unlocked through states outside the set STAB,  hereafter referred to as \textit{magic} states, which allow the implementation of non-stabilizer operations to attain universal fault-tolerant quantum computing.

\subsection{Magic breaking channels}

Given the central role of magic states in universal quantum computation, it is of relevance to characterize channels that can destroy their resource. 

\textbf{Definition $\mathbf{1}$: } {\it Magic-breaking channel}. A channel, $\Lambda^{\mathbb{C}^d \to \mathbb{C}^d}$, is said to be magic-breaking if it transforms {\it all input states into stabilizer states}, i.e., the set of magic-breaking channels may be defined as
\begin{equation}
    \mathcal{M} = \{\Lambda^{\mathbb{C}^d\to\mathbb{C}^d}: \Lambda^{\mathbb{C}^d\to\mathbb{C}^d}(\rho)\in\text{STAB}(\mathbb{C}^d) ~\forall~ \rho \in \mathbb{C}^d \},
    \label{eq:magic-breaking-Cd}
\end{equation}
where the superscript $\mathbb{C}^d\to\mathbb{C}^d$ indicates that the input and output dimensions of the channel, $\Lambda$, are the same.

In the resource theory of entanglement, there exists a convex subset of separable states (free states) that cannot be entangled even with the application of entangling unitaries~\cite{Kraus_PRA_2001, Leifer_PRA_2003, Chefles_PRA_2005}; these are termed absolutely separable states~\cite{Kush_PRA_2001, Verstraete_PRA_2001}. Inspired by this concept, let us introduce another set of magic-breaking channels which is a subset of the aforementioned case.

\textbf{Definition $\mathbf{2}$: } {\it Strictly magic-breaking channel}. A channel is strictly magic-breaking if it outputs stabilizer states which cannot be transformed into non-stabilizer ones even by non-Clifford unitaries. The set of such channels has the form
\begin{eqnarray}
  \nonumber  \tilde{\mathcal{M}} = \{\Lambda^{\mathbb{C}^d\to\mathbb{C}^d}: U_{NC} \circ \Lambda^{\mathbb{C}^d\to\mathbb{C}^d}(\rho)\in\text{STAB}(\mathbb{C}^d) \\
    ~ \forall \rho \in \mathbb{C}^d ~ \text{and} ~ U_{NC} \in \text{non-Clifford unitary operations} \}.
    \label{eq:ab-magic-breaking-Cd}
\end{eqnarray}

\subsection{Properties of magic-breaking channels}
\label{subsec:properties}

Let us establish some ubiquitous properties of magic-breaking channels that are independent of the dimension of the Hilbert space on which the channels act. Note that we only consider finite-dimensional Hilbert spaces in our work.

\textbf{Property $\mathbf{1}$.} \textit{Unitary maps can never be magic-breaking channels.}

\textit{Proof.} In the resource theory of magic states, there are a fixed number of pure stabilizer states in any finite dimension. A unitary map would qualify as magic-breaking if it could transform all pure states into these specific stabilizer states, resulting in a discretized output on the Bloch surface. Since any unitary map keeps the Bloch sphere invariant, it can never be magic-breaking. $\hfill \blacksquare$


\textbf{Property $\mathbf{2}$.} \textit{Channels that destroy the magic of pure states also do so when acting on mixed states.}

\textit{Proof.} Let us assume that a channel, $\Lambda$, renders all pure states into stabilizer states, i.e., $\Lambda (\ket{\eta} \bra{\eta}) \in \text{STAB} ~\forall~ \ket{\eta}$. Since any mixed state can be decomposed into a convex combination of pure states, the action of the considered channel on a mixed state, $\rho$, may be represented as $\Lambda (\rho) = \Lambda (\sum_i p_i \ket{\chi_i} \bra{\chi_i}) = \sum_i p_i \Lambda (\ket{\chi_i} \bra{\chi_i}) \in \text{STAB}$, since each $\Lambda (\ket{\chi_i} \bra{\chi_i}) \in \text{STAB}$ and the stabilizer states form a convex set~\cite{Veitch_NJP_2014}.~$\hfill \blacksquare$

Note that, albeit simple, such a property only holds for channels that output free states belonging to a convex resource theory, e.g., separable states~\cite{Horodecki_RMP_2009} in case of entanglement-breaking channels~\cite{Horodecki_RMP_2003}, non-dense codable states for dense coding breaking channels~\cite{Muhuri_arXiv_2023}, and the like. If the free states do not form a convex set~\cite{Kuroiwa_PRL_2024, Salazar_arXiv_2024}, then the channels destroying the resource in pure states may not be able to do so for mixed states.

\textbf{Property $\mathbf{3}$.} \textit{The magic-breaking channels form a convex and compact set.}

\textit{Proof.} We shall prove this property in two parts.

\textbf{Convexity.} Let $\Lambda_1$ and $\Lambda_2$ be two magic-breaking channels. We immediately have, for $p_1 \in [0,1]$,

\begin{eqnarray}
  \nonumber && (p_1 \Lambda_1 + (1-p_1) \Lambda_2) (\rho) =  p_1 \Lambda_1 (\rho) + (1-p_1) \Lambda_2 (\rho)\\
  && =  p_1 \rho_{\Lambda_1} + (1-p_1) \rho_{\Lambda_2} \in \text{STAB},
\end{eqnarray}
since $\rho_{\Lambda_1}, \rho_{\Lambda_2} \in \text{STAB}$, the last equality follows from the convex structure of the set of stabilizer states. $\hfill \blacksquare$

\textbf{Compactness.} Let us suppose that $\Lambda_0$ is a limit point of the convex set of magic-breaking channels. We construct a sequence of magic-breaking channels, $\{\Lambda_n \in B_n(\Lambda_0)\}$, where each $\Lambda_n$ lies in an open ball, $B_n$, of radius $1/n$ around $\Lambda_0$. By construction, $\lim_{n \to \infty} \Lambda_n \to \Lambda_0$. Further, we consider a second sequence $\{\tau_n = \Lambda_n (\rho)\}$, such that $\lim_{n \to \infty} \tau_n = \tau_0 = \Lambda_0 (\rho)$, i.e., $\tau_0$ is the limit point of $\{\tau_n\}$. Evidently, each $\tau_n \in \text{STAB}$, which, being a polytope, is a closed set~\cite{Gour_arXiv_2024}. Since a closed set contains its limit point, $\tau_0 \in \text{STAB}$ which implies that $\Lambda_0$ is a magic-breaking channel. Thus, the set of magic-breaking channels contains its limit point and is hence compact. $\hfill \blacksquare$

\textbf{Property $\mathbf{4}$.} \textit{Measure-prepare channels, which prepare non-orthogonal stabilizer states, are a class of extreme points for the set of magic-breaking channels.}

\textit{Proof.} Extreme classical-quantum (CQ) channels are measure-prepare channels of the form $\Lambda_{CQ}(\rho) = \sum_{k} \ket{\eta_k} \bra{\eta_k} \langle e_k| \rho|e_k \rangle$, with orthonormal vectors $\{\ket{e_k}\}$. Further, if the prepared states are non-orthogonal, i.e., $\langle \eta_k | \eta_{k'} \rangle \neq 0 ~ \forall ~ k,k'$, the channels are extreme points of the set of CPTP maps~\cite{Horodecki_RMP_2003}. Let us assume that the prepared states are non-orthogonal stabilizer states, making the extreme CQ channels magic-breaking. Thus, such channels are magic-breaking extreme CPTP maps, thereby constituting the extreme points of magic-breaking channels. Hence the proof.~$\hfill \blacksquare$

The set of extreme magic-breaking channels mentioned above is non-unital. It is worth noting that while extreme magic-breaking unital channels do exist, they are neither within the set of extreme unital channels nor are they extreme points of completely positive and trace-preserving channels.

Henceforth, we shall focus only on channels acting on states comprising qubit(s), and examine their magic-breaking properties. 

\section{Magic-breaking qubit channels: criteria and algorithm for identification}
\label{sec:qubit_magic-breaking}

We now provide a characterization of qubit channels capable of breaking magic.
A generic qubit state can be written as $\rho(m_1, m_2, m_3)=\frac{1}{2}(\mathbb{I}_2 + m_1\sigma_1 + m_2\sigma_2 + m_3\sigma_3)$ with the condition $\sum_{i = 1}^3 m_i^2 \leq 1$ on the magnetizations, $m_i = \Tr [\sigma_i \rho(m_1,m_2,m_3)]$, being necessary and sufficient for it to be a valid quantum state~\cite{Preskill}.
Here, $\mathbb{I}_2$ is the two-dimensional identity matrix and $\{\sigma_i\}_{i=1,2,3}$ are the usual Pauli matrices. The state can be represented by a point inside or on the surface of a unit sphere, defined by the axes $\{m_1, m_2, m_3\}$, known as the Bloch sphere. The condition for an arbitrary qubit state, $\rho(m_1, m_2, m_3)$, to be a stabilizer is $|m_1|+|m_2|+|m_3|\leq 1$, which indicates that the qubit stabilizer states lie inside the polytope formed by the axes $\{m_1, m_2, m_3\}$~\cite{Bravyi_PRA_2005}.

An arbitrary qubit channel, $\Lambda$, acting on the two-dimensional complex Hilbert space $\mathbb{C}^2$, can be decomposed as $\Lambda^{\mathbb{C}^2\to\mathbb{C}^2}=U_{\text{post}}\circ \Lambda_C \circ U_{\text{pre}}$~\cite{Ruskai_LAA_2002}, where $U_{\text{pre(post)}}$ is a pre(post)-processing $2\times 2$ unitary matrix and $\Lambda_C$ denotes the canonical form of the qubit channel, whose representation in the Pauli basis $\{\mathbb{I}_2, \sigma_1, \sigma_2, \sigma_3\}$ is

\begin{equation}
\label{eq:canonical_channel}
    \Lambda_C\left(\{t_i\}, \{\lambda_i\}\right)= \begin{pmatrix}
1 & 0 & 0 & 0 \\
t_1 & \lambda_1 & 0 & 0 \\
t_2 & 0 & \lambda_2 & 0 \\
t_3 & 0 & 0 & \lambda_3 
\end{pmatrix}.
\end{equation}
The necessary condition for the complete positivity of $\Lambda_C$ is given by $|t_i|+|\lambda_i|\leq 1~\forall~i$~\cite{Ruskai_LAA_2002}. Note that, Eq.~\eqref{eq:canonical_channel} stands for the most general non-unital channel whereas for a unital channel, which maps the identity state to itself, we have $t_i = 0~\forall~i$.

The action of $\Lambda_C$ on a state $\rho$ reads as $\Lambda_C(\rho)=\frac{1}{2}\left(\mathbb{I}_2 + \sum_{i=1}^3(t_i+\lambda_i m_i)\sigma_i\right)=\frac{1}{2}(\mathbb{I}_2 + m'_1\sigma_1 + m'_2\sigma_2 + m'_3\sigma_3)=\rho(m'_1, m'_2, m'_3)$. Geometrically, the action of $\Lambda_C$ on the Bloch sphere can be manifested as a deformation to a shifted ellipsoid inside the original unit sphere. Note that any deformation of the Bloch sphere into an ellipsoid inside it does not ensure complete positivity~\cite{Ruskai_LAA_2002}. In our study, we presume that the provided map is already a valid CPTP transformation, and subsequently explore its impact on deforming the Bloch sphere. 
Finally, note that any arbitrary $2\times 2$ unitary matrix can be represented as
\begin{equation}
    U(\theta, \phi, \psi)=\begin{pmatrix}
\cos \frac{\theta}{2} ~e^{i(2\pi-\phi-\psi)/2} & i\sin \frac{\theta}{2}~ e^{-i(\phi-\psi)/2} \\
i\sin \frac{\theta}{2}~ e^{i(\phi-\psi)/2} & \cos \frac{\theta}{2} ~e^{-i(2\pi-\phi-\psi)/2}
\end{pmatrix},
\label{eq:unitary}
\end{equation}
whose action on $\mathbb{C}^2$ is isomorphic to a rotation in the $3$-dimensional Euclidean space $\mathbb{R}^3$. More precisely, $U(\theta, \phi, \psi)\rho(m'_1, m'_2, m'_3) U^{\dagger} (\theta, \phi, \psi)=\rho(m''_1, m''_2, m''_3)$, with the state's purity remaining invariant, i.e., ${m'_1}^2 + {m'_2}^2 + {m'_3}^2={m''_1}^2 + {m''_2}^2 + {m''_3}^2$. The pre(post)-processing unitaries can thus be parametrized by $\theta, \phi,$ and $\psi$ according to Eq.~\eqref{eq:unitary}.

To derive conditions on the channel parameters such that the channel is magic-breaking, let us first note that we can dispense with the pre-processing unitary, $U_{\text{pre}}$, since its action on a given state yields another valid state. Therefore, we only need to consider the canonical form, $\Lambda_C(\{t_i\},\{\lambda_i\})$, and the post-processing unitary, $U_{\text{post}}(\theta, \phi, \psi)$, while imposing conditions on $\{t_i, \lambda_k, \theta, \phi, \psi\}$ to characterize qubit magic-breaking channels.

Our analysis of magic-breaking channels rests on the geometric picture of the state space. Given any input to the channel, $\Lambda$, the action of $\Lambda_C$, followed by $U_{\text{post}}$, yields the triad $\mathbf{m''} = \{m''_1, m''_2, m''_3\}$. Note that the action of $U_{\text{post}}$ on the ellipsoid formed by $\Lambda_C$, is to rotate it with respect to the origin within the Bloch sphere while keeping its axis lengths unchanged. If we consider the action of $U_{\text{post}}$ on $\Lambda_C (\rho)$, the final rotated ellipsoid, defined by the axes $\{m''_1, m''_2, m''_3\}$, and containing the output states, is described by $\mu_1^2+\mu_2^2+\mu_3^2=1$, where

\begin{eqnarray}
\label{eq:ellipsoid_parameters-1}
 \mu_1&=&\frac{\sin \psi\left(m''_3\sin\theta + \cos\theta\left(m''_2\cos\phi-m''_1\sin\phi\right)\right)}{\lambda_1}\nonumber\\&+&\frac{\cos\psi\left(m''_1\cos\phi+m''_2\sin\phi\right)}{\lambda_1} - \frac{t_1}{\lambda_1},\\
 \label{eq:ellipsoid_parameters-2}
 \mu_2&=&\frac{m''_3\cos\psi\sin\theta-\sin\psi\left(m''_1\cos\phi+m''_2\sin\phi\right)}{\lambda_2}\nonumber\\&+&\frac{\cos\theta\cos\psi\left(m''_2\cos\phi-m''_1\sin\phi\right)}{\lambda_2} - \frac{t_2}{\lambda_2},\\
 \label{eq:ellipsoid_parameters-3}
 \mu_3&=&\frac{\cos\theta\left(m''_3-m''_2\cos\phi\tan\theta+m''_1\sin\phi\tan\theta\right)}{\lambda_3} \nonumber \\ &-& \frac{t_3}{\lambda_3}.
\end{eqnarray}
Therefore, the above relations allow us to derive the criteria to determine magic-breaking channels in terms of $U(\theta, \phi, \psi)$ and the canonical channel parameters.

\subsection{Algorithm to identify qubit magic-breaking channels}

Let us describe the algorithm to certify magic-breaking channels.
Our algorithm comprises two steps. First, we determine the channel's canonical parameters, $\lambda_i, t_j$, together with the parameters of $U_{\text{post}}$. Then we present a theorem, in terms of the canonical parameters and the variables comprising the post-processing unitary, that determines whether the channel is magic-breaking.

\textbf{Extracting the channel parameters.} Given any complete description of a qubit channel, in terms of its action on a generic state or the same on a qubit basis set, we need to first determine its canonical parameters and the post-processing unitary, $U_{\text{post}}$, in order to check whether it is capable of breaking magic. Given a channel $\Lambda$, we wish to finally arrive at the form $\Lambda = U_{\text{post}}(\theta_2, \phi_2, \psi_2) \circ \Lambda_C(\{t_i\}, \{\lambda_i\}) \circ U_{\text{pre}}(\theta_1, \phi_1, \psi_1)$. We apply this channel on a state, $\rho_{\text{in}}(m_1,m_2,m_3) = \frac{1}{2}(\mathbb{I}_2 + \sum_{j = 1}^{3} m_j \sigma_j)$, and obtain the output state, $\rho_{\text{out}}(m''_1,m''_2,m''_3) = \frac{1}{2}(\mathbb{I}_2 + \sum_{j = 1}^{3} m''_j \sigma_j)$. The output triad is given by $\mathbf{m''} = g(\{m_j\}, \{t_i\}, \{\lambda_i\}, \theta_{1,2}, \phi_{1,2}, \psi_{1,2})$, for some function $g$, which can also be represented as

\begin{eqnarray}
    \vec{m''} = R_2 \Lambda_d R_1 \vec{m} + R_2 \vec{t}.
    \label{eq:channel_determine}
\end{eqnarray}
Here, $R_{1(2)}$ are the rotation matrices corresponding to $U_{\text{pre(post)}}$, $\Lambda_d = \text{diag}(\lambda_1, \lambda_2, \lambda_3)$, $\vec{t} = (t_1, t_2, t_3)^T, \vec{m} = (m_1, m_2, m_3)^T$, and $\vec{m''} = (m''_1, m''_2, m''_3)^T$. The functional form of $\mathbf{m''}$ and Eq.~\eqref{eq:channel_determine} form the data from which the parameters are extracted.

First, we set $\vec{m} = 0$ in the RHS of Eq.~\eqref{eq:channel_determine} to obtain $\vec{m''}_{\text{non-uni}} = R_2 \vec{t}$ which contains the non-unital part of the transformation. The unital transformation is then given as $\vec{m''}_{\text{uni}} = \vec{m''} - \vec{m''}_{\text{non-uni}} = R_2 \Lambda_d R_1 \vec{m}$. Let us represent $R_2 \Lambda_d R_1  = M$. The singular values of $M$ correspond to $\{\lambda_1^2, \lambda_2^2, \lambda_3^2\}$ whereas the diagonalising matrices for $MM^{\dagger}$ and $M^{\dagger} M$ are precisely the matrices $R_2$ and $R_1$, from which we can construct $U_{\text{post}}$ and $U_{\text{pre}}$ respectively, and determine $\{\theta_{1,2}, \phi_{1,2}, \psi_{1,2}\}$ from Eq. \eqref{eq:unitary}. It is now straightforward to determine $\vec{t} = R_2^{-1} \vec{m''}_{\text{non-uni}}$. We have thus extracted all the channel parameters from its action on a generic qubit state. Note that, since we do not consider the action of $U_{\text{pre}}$ in our analysis, we only deal with $\{\theta_2, \phi_2, \psi_2\}$ which we denote as $\{\theta, \phi, \psi\}$ for concision.

We are now equipped to present the main result of our work, which is the necessary and sufficient condition for a qubit channel to be magic-breaking. Given a channel with parameters, $\{\lambda_i, t_j, \theta, \phi, \psi\}$, we first construct the final ellipsoid through Eqs.~\eqref{eq:ellipsoid_parameters-1}-\eqref{eq:ellipsoid_parameters-3}. The following theorem then allows us to determine whether the channel can destroy the magic inherent in any arbitrary single-qubit state.\\
\textbf{Theorem $\mathbf{1}$. \textit{Necessary and sufficient condition for magic-breaking qubit channels}.} {\it The necessary and sufficient criterion for a qubit channel to be magic-breaking is that the final ellipsoid containing the output states must lie within the stabilizer polytope.}

\textit{Proof}. When the final ellipsoid, i.e., $\mu_1^2+\mu_2^2+\mu_3^2=1$, where $\mu_1$, $\mu_2$ and $\mu_3$ are given in Eq. (\ref{eq:ellipsoid_parameters-1})-(\ref{eq:ellipsoid_parameters-3}), lies entirely within the stabilizer polytope (i.e., $|m''_1| + |m''_2| + |m''_3| = 1$), all the output states from the channel also fall inside the stabilizer polytope and are thus stabilizer states. This establishes the sufficiency of the condition. Necessity is dictated by the fact that if any part of the final ellipsoid lies outside the stabilizer polytope, the output states comprising that part are magic states. Therefore, the channel cannot break the magic of all qubit states unless the final ellipsoid lies within the stabilizer polytope. Hence the proof. $\hfill \blacksquare$

\subsubsection*{Analytical condition for magic-breaking qubit channels}

The proof of Theorem $1$ relies on geometrical arguments arising from the action of qubit channels on the Bloch sphere. Let us now derive the analytical conditions that satisfy the above theorem and thus allow us to determine the magic-breaking nature of a given qubit channel. To arrive at the analytical expression corresponding to the theorem, we simultaneously solve the equation for the stabilizer polytope, $|m''_1| + |m''_2| + |m''_3| = 1$, and those for the final ellipsoid, Eqs.~\eqref{eq:ellipsoid_parameters-1} -~\eqref{eq:ellipsoid_parameters-3}, to obtain a condition for no solution or a finite number of solutions. The existence of no solutions for $m''_1$, $m''_2$, and $m''_3$ leads to the fact that either the ellipsoid lies completely inside or outside the polytope. But the complete positivity condition of the quantum channels forbids the latter possibility. When no solutions can be found, it implies that the ellipsoid lies entirely inside the stabilizer polytope whereas a finite number of solutions indicate that the surface of the ellipsoid just touches those of the polytope on the inside, with no intersection between them. {\it Therefore, no or a finite number of simultaneous solutions imply that all the output states lie within the stabilizer polytope and the channel is magic-breaking.} It is worthwhile to mention here, that if some part of the ellipsoid lies outside the polytope, it intersects the latter which leads to an infinite number of solutions for the simultaneous equations. The portion of the final ellipsoid that lies outside the stabilizer polytope contains magic states, which means that the channel cannot destroy the magic of all input states, and, by virtue of Definition $1$, it is not magic-breaking. {\it Hence, the condition of an infinite number of solutions indicates that the channel is not magic-breaking.} Therefore, the necessary and sufficient conditions for magic-breaking are exactly the ones for which the number of simultaneous solutions that exist for all eight faces of the polytope and the rotated ellipsoid is zero or finite.

Let us consider that we simultaneously solve the ellipsoid equation with one face of the stabilizer polytope, say $m''_1 + m''_2 + m''_3 = 1$, to obtain solutions for $m''_2$ and $m''_3$. These solutions are of the form

\begin{eqnarray}
   m''_2 &=& f(m''_1, \{\lambda_i, t_j, \theta, \phi, \psi\}) \pm \sqrt{\alpha {m''_1}^2 + \beta m''_1 + \gamma},~~~~ \label{eq:m2''} \\
      m''_3 &=& f'(m''_1, \{\lambda_i, t_j, \theta, \phi, \psi\}) \mp \sqrt{\alpha {m''_1}^2 + \beta m''_1 + \gamma} ~~~~ \label{eq:m3''}
\end{eqnarray}
The functions $f, ~\text{and}~ f'$ being too complicated and not necessary for our discussion, are not explicitly provided for brevity. Here, $\alpha, \beta, \gamma$ are functions of all the channel parameters and can be determined for a given channel using the aforementioned algorithm. The condition that no or a finite number of solutions exist translates into the expression under the square root being non-positive. If $\alpha {m''_1}^2 + \beta m''_1 + \gamma = 0$, we obtain a finite number of solutions for $m''_1$, which, when substituted into Eqs.~\eqref{eq:m2''} and \eqref{eq:m3''}, yield a finite number of solutions for $m''_2$ and $m''_3$ respectively. This situation corresponds to the final ellipsoid touching the faces of the polytope from the inside. On the other hand, if $\alpha {m''_1}^2 + \beta m''_1 + \gamma < 0$, no real solutions exist for any $m''_i$ which indicates that the ellipsoid lies fully within the stabilizer polytope. However, if $\alpha {m''_1}^2 + \beta m''_1 + \gamma > 0$, there exists an infinite number of simultaneous solutions indicating that a part of the final ellipsoid lies outside the stabilizer polytope, as explained before. Therefore, the necessary and sufficient condition for a qubit channel to be magic-breaking mathematically reads as

\begin{equation}
    \alpha {m''_1}^2 + \beta m''_1 + \gamma \leq 0 .
    \label{eq:necessary-sufficient_math}
\end{equation}
Let us now ask the question -- what relations must $\alpha, \beta, ~\text{and}~ \gamma$ satisfy such that Eq.~\eqref{eq:necessary-sufficient_math} holds true? To gain an insight into the same, we note that $m''_1 \in [-1,1]$ for a valid final state, and that $\alpha {m''_1}^2 + \beta m''_1 + \gamma$ represents a parabola. Depending upon the sign of $\alpha$, the parabola can either be a convex one ($\alpha > 0$) possessing a minimum, or a concave one ($\alpha < 0$) having a maximum. The extremum point in both cases is $ - \frac{\beta}{2 \alpha}$. We shall tackle the two cases separately.\\\\
\textbf{Case $\mathbf{1}: \alpha > 0$.} For a convex parabola to satisfy Eq.~\eqref{eq:necessary-sufficient_math}, the minima must be negative. Consequently, for finite or no solutions to exist for $m''_1 \in [-1,1]$, the parabola must have zeros at or beyond $\pm 1$. Convexity then ensures that Eq.~\eqref{eq:necessary-sufficient_math} is negative at all points within $[-1,1]$. Therefore, no real solutions exist for $m''_i$ and the channel is magic-breaking. This condition occurs when $\alpha \pm \beta + \gamma \leq 0$.\\\\
\textbf{Case $\mathbf{2}: \alpha < 0$.} When the parabola is concave, the maximum can either lie within $[-1,1]$ or beyond this interval, i.e, either $|\frac{\beta}{2 \alpha}| \leq 1$ or $|\frac{\beta}{2 \alpha}| > 1$ respectively. In the first situation, the maximum itself must be non-positive in order to satisfy Eq.~\eqref{eq:necessary-sufficient_math} for all values in $[-1,1]$. This, in turn, translates into $\beta^2 - 4 \alpha \gamma \leq 0$ as the necessary and sufficient condition for magic-breaking. On the other hand, if $|\frac{\beta}{2 \alpha}| > 1$, the maximum can be positive but it lies beyond $\pm 1$. To satisfy Eq.~\eqref{eq:necessary-sufficient_math}, in this case, the parabola must be negative at both $\pm 1$. Therefore, the necessary and sufficient condition for magic-breaking takes the form $\alpha \pm \beta + \gamma \leq 0$.

\textbf{Conditions for magic-breaking.} {\it To summarise, the necessary and sufficient conditions for a qubit channel to be magic-breaking is that the following equations be satisfied for all eight faces of the polytope}
\begin{eqnarray}
\begin{cases}
    \beta^2 - 4 \alpha \gamma \leq 0 & \text{if $\alpha <0$ \text{and} $|\frac{\beta}{4 \alpha}| \leq 1$} \\
    \alpha \pm \beta + \gamma \leq 0 & \text{otherwise}.
\end{cases}
\label{eq:th1_math}
\end{eqnarray}
It is important to note here that for each of the eight faces of the stabilizer polytope, the simultaneous solution with the ellipsoid equation yields a particular set of $\{\alpha, \beta, \gamma\}$, which differs for different faces of the polytope. Magic-breaking is guaranteed if Eq.~\eqref{eq:th1_math} holds true for all the eight sets of $\{\alpha, \beta, \gamma\}$.

\textbf{Note $\mathbf{1}$.} The $T$ state, $\rho_T$, with magnetizations $m_1 = m_2 = m_3 = 1/\sqrt{3}$ is the maximally magic single-qubit state and has robustness of magic (ROM)~\cite{Howard_PRL_2017}, $\mathcal{R}_{\rho_T} = \sqrt{3}$. Since the $T$ state is essential for universal quantum computing but cannot always be prepared with sufficiently high fidelity, it must be distilled from states with lower magic content. It has been established that single-qubit magic states, $\rho$, can be distilled into $T$-type states, using the Bravyi-Kitaev distillation protocol, if their ROM satisfies $\mathcal{R}_{\rho} > 3/\sqrt{7}$~\cite{Bravyi_PRA_2005, Calcluth_PRX_2024}. Since $\mathcal{R}_{\rho(m_1, m_2, m_3)} = \max\{1,~\sum_{i = 1}^{3} |m_i|\}$ for single qubits~\cite{Grismo_PRX_2020}, states lying outside the polytope defined by $\sum_{i = 1}^{3} |m_i| \leq 3/\sqrt{7}$ are surely $T$-distillable. We call this the $T$-polytope. Thus, we can propose the concept of \textit{$T$-distillability breaking channels} which do not completely destroy the magic of the input states, yet impair their distillation into $T$-states, which is similar in spirit to the communication resource-breaking channels analyzed in Ref.~\cite{Muhuri_arXiv_2023}, that are not entanglement-breaking but still prohibit dense coding and teleportation with the output states. \textit{The necessary condition for a qubit channel to be $T$-distillability breaking is that the final ellipsoid containing the output states lies entirely inside the $T$-polytope.} Otherwise, some output states would have $\mathcal{R} > 3/\sqrt{7}$ and would definitely be $T$-distillable. This condition is not sufficient since $\mathcal{R} > 3/\sqrt{7}$ is itself a sufficient condition for $T$-distillability~\cite{Reichardt_QIC_2009, Seddon_PRX_2021} and there may be other distillation procedures involving only Clifford operations that are capable of achieving this. Using our proposed algorithm to extract the channel parameters, we can impose conditions on them such that the channel is necessarily $T$-distillability breaking by simultaneously solving the Eqs. \eqref{eq:ellipsoid_parameters-1} - \eqref{eq:ellipsoid_parameters-3} with $|m_1^{''}| + |m_2^{''}| + |m_3^{''}| = 3/\sqrt{7}$ and obtaining the instances of zero or a finite number of solutions.

\subsection{Different classes of qubit magic-breaking channels}

We now classify qubit-magic-breaking channels into different categories, based on the nature of their post-processing unitary. To begin with, let us derive the condition for magic-breaking irrespective of the involved post-processing operation. This results in a sufficient criterion as follows:

\textbf{Theorem $\mathbf{2}$. \textit{Sufficient criteria for strictly magic-breaking qubit channels.}} \textit{A sufficient condition for a qubit channel to be strictly magic-breaking is $|t|+|\lambda_i|\leq 1/\sqrt{3} ~\forall i$, with $|t|=\sqrt{t_1^2+t_2^2+t_3^2}$.}

\textit{Proof}. Consider the largest sphere inside the stabilizer polytope, with its surface touching the faces of the polytope. The radius of this largest sphere is $1/\sqrt{3}$. 
If the ellipsoid, formed by the canonical channel, $\Lambda_C$, is such that the sum of the distance of its center from the origin and the length of its largest axis becomes smaller than $1/\sqrt{3}$, all the final output states would lie within the largest sphere inside the stabilizer polytope. Any qubit unitary post-processing operation on such states would retain them inside the polytope since such operations cannot enhance the purity of any state. Hence, the sufficient condition for a qubit channel, given in Eq.~\eqref{eq:canonical_channel}, to be magic-breaking is $|t|+|\lambda_i|\leq 1/\sqrt{3} ~\forall i$, with $|t|=\sqrt{t_1^2+t_2^2+t_3^2}$ being the distance between the origin and the center of the ellipsoid.~$\hfill \blacksquare$


It is essential to highlight that the essence of the aforementioned condition lies in its dependence solely on the canonical parameters of a channel. Further this criterion also serves as a sufficient condition for qubit magic-breaking channels.

In the sufficient condition for qubit magic-breaking, thus discussed, we need not deal with the parameters comprising the post-processing unitary. In order to include $U_{\text{post}}$ in our analysis, we must consider specific post-processing operations, since the analysis for arbitrary $\theta, \phi,$ and $\psi$ is mathematically intractable. The necessary and sufficient conditions for different post-processing unitaries can be easily derived from the mathematical expression of Theorem $1$, given in Eq.~\eqref{eq:th1_math}. Let us first consider the simplest post-processing unitary in this scenario, which is a Clifford rotation. 

\textbf{Corollary $\mathbf{1}$. \textit{Qubit channels comprising Clifford post-processing.}} \textit{The necessary and sufficient condition for a qubit channel with Clifford post-processing to be magic-breaking is given by 
\begin{equation}
    \lambda_1^2+\lambda_2^2+\lambda_3^2 \leq \left(1-|t_1|-|t_2|-|t_3|\right)^2.
    \label{eq:cliff_post}
\end{equation}}

\textit{Proof}. The set of stabilizer states remains invariant under Clifford rotations. 
If the canonical component drives the input states into the stabilizer polytope, the Clifford operations ensure that they remain stabilizer states. On the other hand, if the canonical channel does not transform some of the input states into stabilizer states, the Clifford post-processing cannot remove the magic from such states. Therefore, a general qubit channel comprising Clifford post-processing is magic-breaking if and only if its canonical component morphs the Bloch sphere into an ellipsoid inside the stabilizer polytope.  We can thus safely set $\theta = \phi = \psi = 0$ and employ Theorem $1$. In this case, we have $\alpha = - \prod_{i = 1}^{3} \lambda_i^2 \times (\sum_{j = 1}^{3} \lambda_j^2) \leq 0$ and $|\frac{\beta}{2 \alpha}| = \frac{1}{|\sum_{j = 1}^{3} \lambda_j^2|} \times |t_1(\lambda_2^2 + \lambda_3^2) \pm \lambda_1^2(1 - |t_2| - |t_3|)| \leq 1$, whereas $\gamma = \prod_{i = 1}^{3} \lambda_i^2 \Big( \lambda_1^2 (\lambda_2^2 + \lambda_3^2 - (1 - |t_2| - |t_3|)) - t_1^2 (\lambda_2^2 + \lambda_3^2)\Big)$. Thus according to Eq.~\eqref{eq:th1_math}, the necessary and sufficient condition for magic-breaking is given by $\beta^2 - 4 \alpha \gamma \leq 0$ which translates into Eq.~\eqref{eq:cliff_post}. Hence the proof. $\hfill \blacksquare$

Note that Corollary $1$ has an important consequence for Pauli channels. Such channels comprise $\mathcal{I}_2$ as the post-processing operation, which is clearly a Clifford gate (here $\mathcal{I}_2$ is the identity map on qubits, such that $\mathcal{I}_2 (\rho) = \mathbb{I}_2 \rho \mathbb{I}_2)$. \textit{Therefore, Corollary $1$ also provides the necessary and sufficient condition for Pauli channels to be magic-breaking.} 

\textbf{Pauli channels as examples of Corollary $\mathbf{1}$.} \textit{The necessary and sufficient conditions for two paradigmatic Pauli channels to be magic-breaking are as follows: \\
(a) Dephasing channel, $\rho \to (1 - p/2) \rho + (p/2) \sigma_z \rho \sigma_z$ ($0 \leq p \leq 1$)~\cite{Preskill} : $p = 1$,\\
(b) Depolarising channel, $\rho \to p \frac{\mathbb{I}_2}{2} + (1 - p) \rho$ ($0 \leq p \leq 1$)~\cite{Leung_Quantum_2017} :  $p \geq 1-1/\sqrt{3}$.}

\textit{Proof}. Both the channels being unital, we have $\Lambda_C(\{t_i = 0 ~\forall~ i\},\{\lambda_i\})$. For the dephasing channel, $\lambda_1 = \lambda_2 = (1 - p)$, and $\lambda_3 = 1$. Therefore, Eq.~\eqref{eq:cliff_post} translates to $(1 - p)^2 \leq 0$ whose only solution is $p = 1$. This implies that unless the channel is completely dephasing, it cannot be magic-breaking.

On the other hand, one has $\lambda_i = (1 - p) ~\forall~i$ in the case of the depolarising channel. Eq.~\eqref{eq:cliff_post} then translates to $3(1 - p)^2 \leq 1 \implies p \geq 1- 1/\sqrt{3}$. $\hfill \blacksquare$

\textbf{Remark $\mathbf{1}$.} The necessary condition for single-qubit channels having Clifford post-processing to be $T$-distillability breaking is given as $\lambda_1^2 + \lambda_2^2 + \lambda_3^2 \leq (3/\sqrt{7} - |t_1| - |t_2| - |t_3|)^2$, which translates into $p \geq 0.622$, and $p \geq 1 - \sqrt{3/7}$  for the dephasing and the depolarising channels respectively.

\textbf{Entanglement-breaking qubit unital channels comprising Clifford post-processing.} A unital qubit channel is entanglement-breaking (EBT)~\cite{Horodecki_RMP_2003} if and only if its canonical parameters satisfy $\sum_{i = 1}^3 |\lambda_i| \leq 1$~\cite{Ruskai_RMP_2003}. Since, $|\lambda_i|\leq 1 ~\forall i$, $\sum_{j = 1}^3 |\lambda_j| \leq 1\implies\sum_{i=1}^3\lambda_i^2\leq 1$, i.e., the canonical channel is magic-breaking. This suggests that if a qubit unital channel, whose decomposition consists of Clifford post-processing, is entanglement-breaking, it is also a magic-breaking channel. This observation establishes an intriguing relationship between channels that break resources of completely different characters, entanglement being a non-local quantum resource whereas magic is inherently a local feature.

Given any other non-trivial $U_{\text{post}}$, it is possible to characterize \textit{magic-breaking qubit channels} completely by following the prescription to extract the channel parameters and employing Theorem $1$. Due to analytical complexity, we restrict our analysis to a few significant special instances --- the post-processing unitaries corresponding to rotations about the $m''_1, m''_2,$ and $m''_3$ axes. The post-processing unitary, given by Eq.~\eqref{eq:unitary}, corresponding to these rotations is specified by
\begin{eqnarray}
  \nonumber &&  \phi = \psi = 0 ~ \text{for rotation about}~ m''_1, \\
  \nonumber && \phi = -\psi = \pi/2 ~ \text{for rotation about}~ m''_2,~\text{and} \\
  && \theta = \psi = 0~ \text{for rotation about}~ m''_3.
\end{eqnarray}
Substituting these conditions in Eqs.~\eqref{eq:ellipsoid_parameters-1}-\eqref{eq:ellipsoid_parameters-3} and identifying the conditions for finite or no simultaneous solution with the eight polytope faces provides us the instances when the final ellipsoid, after rotation via $U_{\text{post}}$, lies within the stabilizer polytope. For simplicity, we consider that the rotation angle is denoted by $\vartheta$ in all three cases. We thus have

\begin{eqnarray}
   && \nonumber \alpha_i = -\prod_{l = 1}^3 \lambda_l^2 \Big( \sum_{j = 1}^3 \lambda_j^2 + (-1)^{k'} (\lambda_{i \oplus' 1}^2 - \lambda_{i \oplus' 2}^2) \sin 2 \vartheta \Big), \\
   \label{eq:alpha_mi} \\
  && \nonumber |\frac{\beta_i}{2 \alpha_i}| = \frac{|\prod_{l = 1}^3 \lambda_l^2|}{|\alpha_i|} |t_i + (-1)^k \lambda_i^2 \times \\
  && \nonumber \Big\{1 + (-1)^{k \oplus 1} t_i \pm (-1)^{(i-1)k'} \Big(t_{i \oplus' 1} (\cos \vartheta + (-1)^{k'} \sin \vartheta) \\
  && + (-1)^{k'} t_{i \oplus' 2} (\cos \vartheta + (-1)^{k' \oplus 1} \sin \vartheta)\Big)\Big\}|.
\end{eqnarray}
Here, $i = 1, 2, 3$ represent the three axes, $m''_i$, respectively, $k, k' \in \{0,1\}$, $\oplus$ represents addition modulo $2$, and $x \oplus' y$ denotes $(x + y ~\text{modulo}~  2) + 1$. We note that $\alpha_i \leq 0 ~\forall~ i$ but cannot determine the same for $|\frac{\beta_i}{\alpha_i}|$. Eq.~\eqref{eq:th1_math} then allows us to derive the necessary and sufficient conditions for magic-breaking channels comprising post-processing rotations about the three axes in the following: 
\begin{itemize}

\item For $|\frac{\beta_i}{2 \alpha_i}| \leq 1:$ 
\begin{eqnarray}
    && \nonumber \sum_{j = 1}^3 \lambda_j^2 + \text{sgn}(p) \text{sgn}(q) \left(\lambda_{i \oplus' 1}^2-\lambda_{i \oplus' 2}^2\right) \sin 2\vartheta \leq \\
      && \nonumber  \Big(1 + (-1)^k t_i \pm (-1)^{k'+i-1} ( (-1)^{k'} t_{i \oplus' 1} \pm t_{i \oplus' 2})\cos \vartheta \\
      && \mp (-1)^{(k' \oplus 1)+(i-1)} ((-1)^{k' \oplus 1} t_{i \oplus' 2} \pm t_{i \oplus' 1}) \sin \vartheta \Big)^2,
      \label{eq:mi_rot1}
\end{eqnarray}
where $p = t_{i \oplus' 1} \cos \vartheta$, $q = t_{i \oplus' 2} \cos \vartheta$, and sgn() is the sign function.

\item For $|\frac{\beta_i}{2 \alpha_i}| > 1:$
\begin{eqnarray}
&& \nonumber \lambda_i^2 \Big\{ \lambda_{i \oplus' 1}^2 + \lambda_{i \oplus' 2}^2 + (-1)^{k'} (\lambda_{i \oplus' 1}^2 - \lambda_{i \oplus' 2}^2) \sin 2 \vartheta - \\
&& \nonumber (-1)^{(i-1)k'} \Big( t_{i \oplus' 1} (\cos \vartheta + (-1)^{k'} \sin \vartheta) \\
&& \nonumber + t_{i \oplus' 2}((-1)^{k'} \cos \vartheta - \sin \vartheta) \Big)^2 \Big\} \leq \\
&& \nonumber \lambda_{i \oplus' 1}^2 \Big( 1 + (-1)^{k'} \sin 2 \vartheta \Big) \Big( 1 + (-1)^k t_i \Big)^2 \\
&& + \lambda_{i \oplus' 2}^2 \Big( 1 + (-1)^{k' \oplus 1} \sin 2 \vartheta \Big) \Big( 1 + (-1)^k t_i \Big)^2,
\label{eq:mi_rot2}
\end{eqnarray}
and
\begin{eqnarray}
&& \nonumber \lambda_i^2 \Big\{ \lambda_{i \oplus' 1}^2 + \lambda_{i \oplus' 2}^2 + (-1)^{k'} (\lambda_{i \oplus' 1}^2 - \lambda_{i \oplus' 2}^2) \sin 2 \vartheta - \\
&& \nonumber \Big( 2 - (-1)^{(i-1)k' \oplus k}  t_{i \oplus' 1} (\cos \vartheta + (-1)^{k'} \sin \vartheta) \\
&& \nonumber + (-1)^{(i-1)k' \oplus k} t_{i \oplus' 2}((-1)^{k'} \cos \vartheta - \sin \vartheta) \Big)^2 \Big\} \leq \\
&& \nonumber \lambda_{i \oplus' 1}^2 \Big( 1 + (-1)^{k'} \sin 2 \vartheta \Big) \Big( 1 + (-1)^{k''} t_i \Big)^2 \\
&& + \lambda_{i \oplus' 2}^2 \Big( 1 + (-1)^{k' \oplus 1} \sin 2 \vartheta \Big) \Big( 1 + (-1)^{k''} t_i \Big)^2,
\label{eq:mi_rot3}
\end{eqnarray}
where, $k^{''} \in \{0, 1\}$.
\end{itemize}

It can be readily observed that if the rotation angle vanishes, all the conditions mentioned above simplify to that of qubit channels composed of $\Lambda_C$ and Clifford post-processing. We can now present our next corollary to Theorem $1$.

\textbf{Corollary $\mathbf{2}$.} \textit{The necessary and sufficient conditions for qubit channels to be magic-breaking for post-processing unitaries corresponding to rotations about $m''_i$ axes, are given by Eqs.~\eqref{eq:mi_rot1} -~\eqref{eq:mi_rot3} for each realization of $k, k^{'},$ and $k^{''}$.}

\textbf{Remark $\mathbf{2}$.} In the aforementioned discussion, assigning $t_{j=1,2,3}=0$ yields the similar kind of results for magic-breaking unital qubit channels. In Appendix \ref{app:unital}, we provide the necessary and sufficient condition for any unital channel, comprising generic post-processing operations, to be magic-breaking.
\section{Multi-qubit magic-breaking channels: necessary and sufficient conditions}
\label{sec:multi-qubit_MB}

We now turn our focus on channels that can destroy the magic of multi-qubit states. We note that magic, unlike non-local correlations such as entanglement, is a local property. Operationally, it quantifies the advantage leveraged from quantum states for fault-tolerant quantum computing at a single location where the quantum computer is in use. Therefore, contrary to channels breaking quantum resources required for non-local quantum processes, e.g., entanglement-breaking~\cite{Horodecki_RMP_2003} channels or communication resource-breaking channels~\cite{Muhuri_arXiv_2023}, multi-qubit magic-breaking channels act on the entire state instead of on a subsystem. Specifically, we shall consider the tensor product of $N$ single-qubit channels to act on an $N$-qubit state. 

\textbf{Theorem $\mathbf{4}$. \textit{Necessary condition}.} \textit{The necessary condition for $N$ local qubit channels acting on an $N$-qubit state to be magic-breaking is when each channel is magic-breaking.}

\textit{Proof.} We shall prove this statement by contradiction. Let us consider that the tensor product channel $\Lambda_1 \otimes \cdots \otimes \Lambda_N$ acting on an $N$-qubit state, $\rho_N$, is magic-breaking, but the constituent qubit channels, $\Lambda_{i}$, are not. Then, the subsystem corresponding to the qubit $x$ is given by $\Tr_{\bar{x}}(\Lambda_1 \otimes \cdots \otimes \Lambda_N (\rho_N)) = \Lambda_x (\rho_x)$, where $\bar{x}$ denotes all other parties except $x$, and $\rho_x=\Tr_{\bar{x}}(\rho_N)$. If $\Lambda_x$ is not a magic-breaking channel, $\Lambda_x (\rho_x)$ may be a magic state for some initial state $\rho_N$. But since we consider the tensor product channel to be magic-breaking, $(\Lambda_1 \otimes \cdots \otimes \Lambda_N)(\rho_N) \in \text{STAB} ~\forall~ \rho_N$, which implies that one can obtain a magic state from a stabilizer state through partial tracing, a free operation in the resource theory of magic~\cite{Veitch_NJP_2014}. Therefore, the situation described above is not possible, and for $\Lambda_1 \otimes \cdots \otimes \Lambda_N$ to be magic-breaking, each $\Lambda_x$ must individually be magic-breaking. Hence the proof. $\hfill \blacksquare$

\textbf{Remark $\mathbf{3}$.} Recall that we demonstrated the magic-breaking capability of a qubit unital entanglement-breaking channel, $\Lambda_{\text{EBT}}$, when its decomposition consists of Clifford post-processing. Notably, $(\Lambda_{\text{EBT}} \otimes \mathcal{I}_2)$ can destroy the entanglement of any two-qubit entangled state by acting locally on a subsystem. However, an immediate consequence of Theorem $4$ is that $(\Lambda_{\text{EBT}} \otimes \mathcal{I}_2)$ cannot be a two-qubit magic-breaking channel. We thus establish further vindication of the inherent local character of magic as a resource, due to which the local qubit channels must act on each individual qubit in order for the tensor product channel to be magic-breaking. 

A natural question arises whether the necessary condition established herein also serves as a sufficient condition. Our numerical studies confirm that the answer is negative, which allows us to present the following observation:

\textit{\textbf{Observation $\mathbf{1}$:} The tensor product of multiple single-qubit magic-breaking channels may not be magic-breaking for a multi-qubit state.}

The statement can be demonstrated through an example with two qubits. Consider a single-qubit unital magic breaking channel characterized by the canonical component, $\Lambda_C = \text{diag}(1, \lambda_1 = -0.9, \lambda_2 = -0.3, \lambda_3 = 0.2$), and comprising Clifford post-processing. The action of $\Lambda_C^{\otimes 2}$ on the two-qubit pure state, $\ket{\eta} = ( - 0.482 - \iota 0.648) \ket{00} + (0.015 - 0.022 \iota) \ket{01} + (-0.131 - 0.098 \iota) \ket{10} + ( - 0.145 - 0.548 \iota) \ket{11}$ (here, $\iota = \sqrt{-1})$, having initial robustness of magic (ROM)~\cite{Howard_PRL_2017}, $\mathcal{R}_{\ket{\eta}} = 1.834$, is to reduce the magic content to $\mathcal{R}_{\Lambda_C^{\otimes 2}( \ket{\eta})} = 1.0212$. Since $\mathcal{R} > 1$ iff the state is non-stabilizer~\cite{Howard_PRL_2017}, the aforementioned channel cannot break the magic of the state, $\ket{\eta}$, despite both the constituent channels being magic-breaking for single-qubit states $(\sum_{i = 1}^3 \lambda_i^2 = 0.94 < 1)$, according to Corollary $1$.

The above result can easily be extended to an arbitrary number of qubits, which leaves open the question of a sufficient condition for local channels, acting on different parties, to be magic-breaking. 

\subsection{Implications of Observation $1$}

The insufficiency of the tensor product of two single-qubit magic-breaking channels to be magic-breaking for generic two-qubit states has extremely important consequences.

\textbf{$\mathbf{1}$. Single-qubit magic-breaking channels acting locally on individual subsystems of a multi-qubit fully separable state, which is completely product in all bipartitions, can always break the magic of the state.} Let us consider the $N$-qubit state, $\rho_N = \sum_i p_i \rho_i^1 \otimes \cdots \otimes \rho_i^N$, which is product in all bipartitions with $\rho_i^j$ denoting the subsystem, $j$, of the state $i$ in the convex mixture. Clearly, $\otimes_{j = 1}^N \Lambda_j (\rho_N)$ is a stabilizer state if all $\Lambda_j$ are magic breaking, since $\Lambda_j (\rho_i^j) \in \text{STAB} ~\forall~ i \implies \otimes_{j = 1}^N \Lambda_j(\rho_i^1 \otimes \cdots \otimes \rho_i^N) \in \text{STAB}$ which forms a convex set. Together with Observation $1$, this implies that although the tensor product of magic-breaking channels can render individual subsystems into stabilizer states, they may not always be able to destroy the magic present in the correlations. Prime examples of states that possess magic only in their correlations are some two-qubit maximally entangled states, whose local subsystems are the maximally mixed state, $\mathbb{I}_2/2$. This discussion suggests that channels acting globally on all parties of a multi-qubit correlated state may be needed to sufficiently break its magic.

As an immediate result of the above discussion is that the conjugation of entanglement-breaking~\cite{Horodecki_RMP_2003, Ruskai_RMP_2003} and magic-breaking channels can always be magic-breaking, i.e., $\otimes_{j = 1}^N \Lambda_i (\rho_N) \circ \otimes_{j = 1}^{N-1} \Lambda_{\text{EBT}_j}$ is capable of destroying the magic of any $N$-qubit state, since $\otimes_{j = 1}^{N-1} \Lambda_{\text{EBT}_i}$ ensures that its output is a fully separable state, comprising a convex mixture of states product in all bipartitions. This is intriguing since neither type of channel may be $N$-qubit magic-breaking on their own. Moreover, such magic-breaking channels can also be constructed out of $N$ qubit entanglement-annihilating channels~\cite{Filipov_PRA_2012, Filipov_PRA_2013, Filippov_PRA_2013_2}, which act on all the parties, before the application of the individual magic-breaking channels.

\textbf{$\mathbf{2}$. Foundation for the dynamical resource theory of magic preservability.} In the static resource theory of magic, the free states are the resourceless stabilizer states, while free channels comprise stabilizer operations. The tensor product of stabilizer states being also a stabilizer state, there is no scope for resource superactivation~\cite{Masanes_PRL_2008, Palazuelos_PRL_2012, Liang_PRA_2012, Quintino_PRA_2016, Hsieh_PRA_2016}, unlike the resource theories of non-locality~\cite{Vicente_JPA_2014, Brunner_RMP_2014} and steering~\cite{Gallego_PRX_2015, Uola_RMP_2020}. This implies that stabilizer states are in fact \textit{absolutely free states}~\cite{Hsieh_Quantum_2020} in the static resource theory of magic, similar to separable states in the static resource theory of entanglement~\cite{Vidal_PRL_1997, Horodecki_RMP_2009}.

Since quantum channels are inevitable for any information-processing task, it is of fundamental importance to identify channels that can preserve the resource inherent in a resource state~\cite{Hsieh_Quantum_2020}. In such dynamical resource theories induced by the corresponding static resource theory (c.f.~\cite{Stratton_PRL_2024} for the case of purity), the free channels are the ones that completely destroy the resource, which, in our case, are the single-qubit magic-breaking channels. However, despite there being no superactivation of the resource itself, Observation $1$ indicates that there may be activation of resource preservability~\cite{Hsieh_PRA_2016}. This leads to the concept of \textit{absolutely magic-breaking channels}, i.e., magic-breaking channels $\tilde{\Lambda}$ such that $\tilde{\Lambda} \otimes \Lambda$ is magic-breaking for all generic magic-breaking channels, $\Lambda$, thereby preventing activation of magic preservability in those channels. Note that the completely depolarising qubit channel, $\Lambda_{\text{depo}} (\rho) = \Tr(\rho) \frac{\mathbb{I}}{2}$ is an example of qubit absolutely magic-breaking channels. Identification of such channels is of paramount importance since they possess the ability to impair universal quantum computation by destroying the magic present even in the correlations. On a related note, local operations and classical communication (LOCC)~\cite{Chitambar_CMP_2014} which are both entanglement-breaking and entanglement-annihilating~\cite{Moravcikova_JPA_2010} are absolutely entanglement-destroying channels~\cite{Hsieh_Quantum_2020}.

The discussion above motivates us to define free superchannels~\cite{Gour_arXiv_2020} in the dynamical resource theory of magic preservability. The free superchannels take stabilizer operations to stabilizer ones, while forbidding the creation of any magic preservability, and are defined as

\begin{eqnarray}
    \Theta (\mathcal{E}) = \mathcal{P} \circ (\mathcal{E} \otimes \tilde{\Lambda}) \circ \mathcal{Q},
    \label{eq:free-superchannel}
\end{eqnarray}
where $\mathcal{P}$ and $\mathcal{Q}$ are pre- and post-processing stabilizer operations, while the absolutely magic-breaking channel, $\tilde{\Lambda}$ acts on an auxiliary system~\cite{Hsieh_Quantum_2020}. Note that if $\mathcal{E}$ is a magic-breaking channel, $\Theta (\mathcal{E})$ is also magic-breaking. It will be interesting to further devise 
distance-based dynamical resource monotones~\cite{Gour_arXiv_2020} which would allow us to characterize the interconversion between stabilizer operations in terms of magic-preservability.

\section{Conclusion}
\label{sec:conclu}

The quantum world is shown to offer certain advantages in communication and computational abilities over its classical counterpart. In this context, significant studies have been conducted to identify channels, fundamentally required to transmit information, that can destroy specific quantum resources such as entanglement, coherence, and nonlocality. Recently, these studies have shifted towards an operational viewpoint, considering information-theoretic process-breaking channels, which can make resourceful states useless for a fixed quantum protocol. Given that magic is the cornerstone for universal quantum computation, our present study concerns channels that can eliminate the resource of magic from states necessary for the same.


To summarize our findings, we enumerated and proved the properties of magic-breaking channels in arbitrary dimensions.~Demonstrating that magic-breaking channels constitute a convex and compact set, we partially characterized the extreme points within this set. Subsequently, we proposed an algorithm for discerning whether a qubit channel is magic-breaking. Furthermore, we delved into various classes of magic-breaking channels, providing a sufficient criteria in the qubit regime. The necessary and sufficient conditions for qubit magic-breaking channels were also derived and examples of magic-breaking channels under specific post-processing unitary operations were provided. Additionally, extending the consideration to multi-qubit magic-breaking channels, we proved that a tensor product of local qubit magic-breaking channels is necessarily a multi-qubit magic-breaking channel, although this criterion is not sufficient. We also established a relationship between entanglement-breaking and magic-breaking channels, both in the single-qubit as well as in the multi-qubit regimes.

This study lays the groundwork for advancing the dynamical resource theory of magic beyond its static framework. Here, magic-breaking channels may be treated as free entities, while superchannels comprising stabilizer operations are deemed as free operations. Given the convex and compact nature of the set of magic-breaking channels, distance-based metrics can be utilized to measure the dynamical resource of magic. Our work raises several open questions: Given that a tensor product of magic-breaking channels is not always magic-breaking, can we fully characterize absolutely magic-breaking channels that will always prevent the activation of resource preservability? Based on the properties of such channels, it will be interesting to formulate the dynamical resource theory of magic preservability. 

From an applicative standpoint, it would be interesting to explore the interplay between magic-breaking channels and quantum computers, particularly in contexts where communication plays a crucial role in the computation process. One relevant scenario is modular or distributed quantum computing~\cite{monroe2014large, van2016path}, where multiple quantum processors, each handling partial information processing tasks, are quantum-coherently linked to accomplish a specific objective. In such systems, the quality of inter-processor links can impact the integrity of the information transmitted, potentially degrading the overall computational performance. Analyzing magic-breaking phenomena in these links could thus establish fundamental limits on the performance of the distributed computation. Another potentially relevant context is blind quantum computing~\cite{fitzsimons2017private}. Here, quantum computation is outsourced by a client to a server while maintaining the privacy of both input and output data. This process typically involves a noisy quantum channel between the client and the server~\cite{takeuchi2016blind, sheng2018blind}. Analyzing the magic-breaking properties of such channels in specific settings may provide valuable insights into the computational capabilities and limitations of this delegated quantum computing approach as well.

\section*{Acknowledgement}

A.P., R.G., and A.S.D. acknowledge the support from the Interdisciplinary Cyber-Physical Systems (ICPS) program of the Department of Science and Technology (DST), India, Grant No.: DST/ICPS/QuST/Theme- $1/2019/23$. This research was supported in part by the ``INFOSYS scholarship for senior students''. A.F. and R.G. acknowledge funding from the HORIZON-EIC-$2022$-PATHFINDERCHALLENGES-$01$ program under Grant Agreement No.~$10111489$ (Veriqub). Views and opinions expressed are however those of the authors only and do not necessarily reflect those of the European Union. Neither the European Union nor the granting authority can be held responsible for them. \\

\appendix

\section{The Weyl-Heisenberg operators}
\label{subsubsec:stab}

Given a prime dimension, $d$, we can define the generalized Pauli operators (or the Weyl-Heisenberg operators) as

\begin{eqnarray}
    \mathcal{T}_{k,l} = \begin{cases}
        \iota^{kl} Z^k X^l ~~ \text{for}~~ d = 2 \\
        \omega^{-\frac{kl}{2}} Z^k X^l ~~ \text{for}~~ d > 2,
    \end{cases}
\end{eqnarray}
where $\{k,l\} \in \{0, \cdots, d-1\}, \iota = \sqrt{-1}, \omega = \exp(2 \pi \iota/d)$, and the operators $X$ and $Z$ are defined by their action of the computational basis as $X \ket{n} = \ket{n\oplus1}, Z \ket{n} = \omega^n \ket{n}$ with $\oplus$ representing addition modulo $d$~\cite{Veitch_NJP_2014}.
\begin{widetext}

\section{Magic-breaking qubit unital channels}
\label{app:unital}

We use Theorem $1$ to derive the necessary and sufficient conditions for a generic unital channel to be magic-breaking. Unfortunately, there is no closed-form expression for the same. In this section, we solve the final ellipsoid equation with all eight faces of the polytope and find the eight values of $\alpha, \beta, ~\text{and}~ \gamma$, which must satisfy Eq.~\eqref{eq:th1_math}. We first enumerate the variables $\alpha_i$ below:

\begin{eqnarray}
\nonumber && \alpha_1 = \alpha_8 = \\
\nonumber && 4 \left(\lambda_1^2+\lambda_2^2+\lambda_3^2\right) +  4 \sin \theta \Big(\lambda_1^2+\lambda_2^2-2 \lambda_3^2 \Big) \Big( \sin \theta \sin \phi \cos \phi + \cos \theta (\cos \phi-\sin \phi) \Big) \\ 
\nonumber && -(\lambda_1^2-\lambda_2^2) \cos 2 \psi \Big(2 \sin 2 \theta  (\cos \phi-\sin \phi) -(3 + \cos 2 \theta) \sin 2 \phi \Big) \\
&& \nonumber + 4 (\lambda_1^2-\lambda_2^2) \sin 2 \psi (\sin \phi+\cos \phi) \Big(\sin \theta  + \cos \theta (\cos \phi-\sin \phi)\Big), \\
\end{eqnarray}

\begin{eqnarray}
\nonumber && \alpha_2 = \alpha_7 = \\
\nonumber && 4 \left(\lambda_1^2+\lambda_2^2+\lambda_3^2\right) +  4 \sin \theta \Big(\lambda_1^2+\lambda_2^2-2 \lambda_3^2 \Big) \Big(\sin \theta \sin \phi \cos \phi - \cos \theta (\cos \phi-\sin \phi) \Big) \\
\nonumber && +(\lambda_1^2-\lambda_2^2) \cos 2 \psi \Big(2 \sin 2 \theta  (\cos \phi-\sin \phi) +(3 + \cos 2 \theta) \sin 2 \phi \Big) \\
&& \nonumber +4 (\lambda_1^2-\lambda_2^2) \sin 2 \psi (\sin \phi+\cos \phi) \Big(\sin \theta - \cos \theta (\cos \phi-\sin \phi) \Big), \\
\end{eqnarray}

\begin{eqnarray}
\nonumber && \alpha_3 = \alpha_6 = \\
\nonumber && 4 \left(\lambda_1^2+\lambda_2^2+\lambda_3^2\right) -  4 \sin \theta \Big(\lambda_1^2+\lambda_2^2-2 \lambda_3^2 \Big) \Big( \sin \theta \sin \phi \cos \phi + \cos \theta (\cos \phi+\sin \phi) \Big) \\ 
\nonumber && -(\lambda_1^2-\lambda_2^2) \cos 2 \psi \Big(2 \sin 2 \theta  (\cos \phi+\sin \phi) -(3 + \cos 2 \theta) \sin 2 \phi \Big) \\
&& \nonumber + 4 (\lambda_1^2-\lambda_2^2) \sin 2 \psi (\cos \phi-\sin \phi) \Big(\sin \theta  - \cos \theta (\cos \phi+\sin \phi)\Big), \\
\end{eqnarray}

\begin{eqnarray}
\nonumber && \alpha_4 = \alpha_5 = \\
\nonumber && 4 \left(\lambda_1^2+\lambda_2^2+\lambda_3^2\right) -  4 \sin \theta \Big(\lambda_1^2+\lambda_2^2-2 \lambda_3^2 \Big) \Big( \sin \theta \sin \phi \cos \phi - \cos \theta (\cos \phi+\sin \phi) \Big) \\ 
\nonumber && -(\lambda_1^2-\lambda_2^2) \cos 2 \psi \Big(2 \sin 2 \theta  (\cos \phi+\sin \phi) +(3 + \cos 2 \theta) \sin 2 \phi \Big) \\
&& \nonumber - 4 (\lambda_1^2-\lambda_2^2) \sin 2 \psi (\cos \phi-\sin \phi) \Big(\sin \theta  + \cos \theta (\cos \phi+\sin \phi)\Big). \\
\end{eqnarray}

Next, let us provide the expressions for $\beta_i$

\begin{eqnarray}
  \nonumber && \beta_1 = - \beta_8 = \lambda_1^2 \lambda_2^2 \lambda_3^2 \times \\
 \nonumber  &&   \Big( \Big(3 \lambda_1^2 + 3 \lambda_2^2 + 2 \lambda_3^2 \Big) +  \left( \lambda_1^2+\lambda_2^2-2 \lambda_3^2\right) \left( \cos 2 \theta -2 \sin 2\theta \sin \phi+2 \sin ^2\theta (\sin 2 \phi+\cos 2 \phi)  \right)\\ 
\nonumber && + (\lambda_1^2-\lambda_2^2) \cos 2 \psi \left(2 \sin 2\theta \sin \phi + (3 + \cos 2 \theta) \cos 2 \phi+(3 + \cos 2 \theta) \sin 2 \phi+2 \sin ^2\theta\right) \\
 && +4 (\lambda_1^2-\lambda_2^2) \sin 2 \psi \Big(\sin \theta \cos \phi + \cos \theta (\cos 2 \phi-\sin 2 \phi) \Big)\Big),
\end{eqnarray}

\begin{eqnarray}
  \nonumber && \beta_2 = - \beta_7 = \lambda_1^2 \lambda_2^2 \lambda_3^2 \times \\
 \nonumber  && \Big( \Big(3 \lambda_1^2 + 3 \lambda_2^2 + 2 \lambda_3^2 \Big) + \left( \lambda_1^2+\lambda_2^2-2 \lambda_3^2\right) \left( \cos 2 \theta +2 \sin 2\theta \sin \phi+2 \sin ^2\theta (\sin 2 \phi+\cos 2 \phi)  \right)\\ 
\nonumber && + (\lambda_1^2-\lambda_2^2) \cos 2 \psi \left( - 2 \sin 2\theta \sin \phi + (3 + \cos 2 \theta) \cos 2 \phi+(3 + \cos 2 \theta) \sin 2 \phi+2 \sin ^2\theta\right) \\
 && +4 (\lambda_1^2-\lambda_2^2) \sin 2 \psi \Big(\sin \theta \cos \phi - \cos \theta (\cos 2 \phi-\sin 2 \phi) \Big)\Big),
\end{eqnarray}

\begin{eqnarray}
  \nonumber && \beta_3 = - \beta_6 = \lambda_1^2 \lambda_2^2 \lambda_3^2 \times \\
 \nonumber  &&   \Big( \Big(3 \lambda_1^2 + 3 \lambda_2^2 + 2 \lambda_3^2 \Big) +  \left( \lambda_1^2+\lambda_2^2-2 \lambda_3^2\right) \left( \cos 2 \theta -2 \sin 2\theta \sin \phi+2 \sin ^2\theta ( \cos 2 \phi - \sin 2 \phi)  \right)\\ 
\nonumber && + (\lambda_1^2-\lambda_2^2) \cos 2 \psi \left(2 \sin 2\theta \sin \phi + (3 + \cos 2 \theta) \cos 2 \phi - (3 + \cos 2 \theta) \sin 2 \phi+2 \sin ^2\theta\right) \\
 && +4 (\lambda_1^2-\lambda_2^2) \sin 2 \psi \Big(\sin \theta \cos \phi - \cos \theta (\cos 2 \phi+\sin 2 \phi) \Big)\Big),
\end{eqnarray}

\begin{eqnarray}
  \nonumber && \beta_4 = - \beta_5 = \lambda_1^2 \lambda_2^2 \lambda_3^2 \times \\
 \nonumber  &&   \Big( \Big(3 \lambda_1^2 + 3 \lambda_2^2 + 2 \lambda_3^2 \Big) +  \left( \lambda_1^2+\lambda_2^2-2 \lambda_3^2\right) \left( \cos 2 \theta +2 \sin 2\theta \sin \phi+2 \sin ^2\theta (\cos 2 \phi - \sin 2 \phi)  \right)\\ 
\nonumber && + (\lambda_1^2-\lambda_2^2) \cos 2 \psi \left(-2 \sin 2\theta \sin \phi + (3 + \cos 2 \theta) \cos 2 \phi-(3 + \cos 2 \theta) \sin 2 \phi+2 \sin ^2\theta\right) \\
 && +4 (\lambda_1^2-\lambda_2^2) \sin 2 \psi \Big(\sin \theta \cos \phi + \cos \theta (\cos 2 \phi+\sin 2 \phi) \Big)\Big).
\end{eqnarray}

Finally, the expressions for $\gamma_i$ read as

\begin{eqnarray}
   && \nonumber \gamma_1 = \gamma_4 = \gamma_5 = \gamma_8 = 4 \lambda_1^2 \lambda_2^2 \lambda_3^2 \times \\
   && \nonumber \Big(\cos ^2\theta \left(\frac{1}{2} \cos ^2\phi \left(\cos 2 \theta+\lambda_3^2-1\right) \left(\lambda_1^2+(\lambda_1^2-\lambda_2^2) \cos 2 \psi+\lambda_2^2\right)-\sin ^2\phi \left(\lambda_1^2 \sin ^2\psi+\lambda_2^2 \cos ^2\psi\right)+\lambda_1^2 \lambda_2^2\right) \\
   && \nonumber +\frac{1}{2} \cos \theta \cos \phi \left(-4 \lambda_3^2 \sin \theta \left(\lambda_1^2 \cos ^2\psi+\lambda_2^2 \sin ^2\psi\right)-(\lambda_1^2-\lambda_2^2) \sin 2 \psi \sin \phi \left(\cos 2 \theta+2 \lambda_3^2-1\right)\right) \\
   && \nonumber +\lambda_3^2 \Big(\sin ^2\psi \left(\sin ^2\theta \left(\lambda_2^2-\sin ^2\psi \sin ^2\phi\right)+\lambda_1^2 \sin ^2\phi\right)+\cos ^2\psi \left(\sin ^2\theta \left(\lambda_1^2-2 \sin ^2\psi \sin ^2\phi\right)+\lambda_2^2 \sin ^2\phi\right) \\
   && \nonumber +\lambda_1^2 \sin \theta \sin 2 \psi \sin \phi-2 \lambda_2^2 \sin \theta \sin \psi \cos \psi \sin \phi-\sin ^2\theta \cos ^4\psi \sin ^2\phi\Big)\\
   && \nonumber +\cos ^2\phi \left(\lambda_1^2 \lambda_2^2 \sin ^2\theta-\sin ^4\theta \left(\lambda_1^2 \cos ^2\psi+\lambda_2^2 \sin ^2\psi\right)\right) - \cos ^4\theta \left(\cos ^2\phi\right) \left(\lambda_1^2 \cos ^2\psi+\lambda_2^2 \sin ^2\psi\right)+\lambda_1^2 \lambda_2^2 \sin 2 \theta \cos \phi\\
   && +2 \cos ^3\theta (\lambda_1-\lambda_2) (\lambda_1+\lambda_2) \sin \psi \cos \psi \sin \phi \cos \phi \Big),
\end{eqnarray}

\begin{eqnarray}
   && \nonumber \gamma_2 = \gamma_3 = \gamma_6 = \gamma_7 = 4 \lambda_1^2 \lambda_2^2 \lambda_3^2 \times \\
   && \nonumber \Big( \cos ^2\theta \left(\frac{1}{2} \cos ^2\phi \left(\cos 2 \theta+\lambda_3^2-1\right) \left(\lambda_1^2+(\lambda_1^2-\lambda_2^2) \cos 2 \psi+\lambda_2^2\right)-\sin ^2\phi \left(\lambda_1^2 \sin ^2\psi+\lambda_2^2 \cos ^2\psi\right)+\lambda_1^2 \lambda_2^2\right) \\
   && \nonumber +\lambda_3^2 \sin 2 \theta \cos \phi \left(\lambda_1^2 \cos ^2\psi+\lambda_2^2 \sin ^2\psi\right)+\lambda_3^2 \Big(\sin ^2\psi \left(\sin ^2\theta \left(\lambda_2^2-\sin ^2\psi \sin ^2\phi\right)+\lambda_1^2 \sin ^2\phi\right) \\
   && \nonumber +\cos ^2\psi \left(\sin ^2\theta \left(\lambda_1^2-2 \sin ^2\psi \sin ^2\phi\right)+\lambda_2^2 \sin ^2\phi\right)-2 \lambda_1^2 \sin \theta \sin \psi \cos \psi \sin \phi+\lambda_2^2 \sin \theta \sin 2 \psi \sin \phi-\sin ^2\theta \cos ^4\psi \sin ^2\phi\Big) \\
   && \nonumber +\cos ^2\phi \left(\lambda_1^2 \lambda_2^2 \sin ^2\theta-\sin ^4\theta \left(\lambda_1^2 \cos ^2\psi+\lambda_2^2 \sin ^2\psi\right)\right) - \cos ^4\theta \left(\cos ^2\phi\right) \left(\lambda_1^2 \cos ^2\psi+\lambda_2^2 \sin ^2\psi\right)-\lambda_1^2 \lambda_2^2 \sin 2 \theta \cos \phi \\
   && \nonumber +2 \cos \theta \cos \phi ((\lambda_1^2-\lambda_2^2)  \sin \psi \cos \psi \sin \phi (\sin \theta-\lambda_3) (\sin \theta+\lambda_3)) \\
   && +2 \cos ^3\theta (\lambda_1^2-\lambda_2^2) \sin \psi \cos \psi \sin \phi \cos \phi \Big).
\end{eqnarray}

All unital channels which satisfy Eq.~\eqref{eq:th1_math} $\forall~ i = 1, 2, \cdots, 8$ are guaranteed to be magic-breaking channels, since the final ellipsoid then always lies inside the stabilizer polytope.

\end{widetext}

\bibliographystyle{apsrev4-1}
	\bibliography{ref}

\begin{thebibliography}{114}%
\makeatletter
\providecommand \@ifxundefined [1]{%
 \@ifx{#1\undefined}
}%
\providecommand \@ifnum [1]{%
 \ifnum #1\expandafter \@firstoftwo
 \else \expandafter \@secondoftwo
 \fi
}%
\providecommand \@ifx [1]{%
 \ifx #1\expandafter \@firstoftwo
 \else \expandafter \@secondoftwo
 \fi
}%
\providecommand \natexlab [1]{#1}%
\providecommand \enquote  [1]{``#1''}%
\providecommand \bibnamefont  [1]{#1}%
\providecommand \bibfnamefont [1]{#1}%
\providecommand \citenamefont [1]{#1}%
\providecommand \href@noop [0]{\@secondoftwo}%
\providecommand \href [0]{\begingroup \@sanitize@url \@href}%
\providecommand \@href[1]{\@@startlink{#1}\@@href}%
\providecommand \@@href[1]{\endgroup#1\@@endlink}%
\providecommand \@sanitize@url [0]{\catcode `\\12\catcode `\$12\catcode `\&12\catcode `\#12\catcode `\^12\catcode `\_12\catcode `\%12\relax}%
\providecommand \@@startlink[1]{}%
\providecommand \@@endlink[0]{}%
\providecommand \url  [0]{\begingroup\@sanitize@url \@url }%
\providecommand \@url [1]{\endgroup\@href {#1}{\urlprefix }}%
\providecommand \urlprefix  [0]{URL }%
\providecommand \Eprint [0]{\href }%
\providecommand \doibase [0]{http://dx.doi.org/}%
\providecommand \selectlanguage [0]{\@gobble}%
\providecommand \bibinfo  [0]{\@secondoftwo}%
\providecommand \bibfield  [0]{\@secondoftwo}%
\providecommand \translation [1]{[#1]}%
\providecommand \BibitemOpen [0]{}%
\providecommand \bibitemStop [0]{}%
\providecommand \bibitemNoStop [0]{.\EOS\space}%
\providecommand \EOS [0]{\spacefactor3000\relax}%
\providecommand \BibitemShut  [1]{\csname bibitem#1\endcsname}%
\let\auto@bib@innerbib\@empty
\bibitem [{\citenamefont {Nielsen}\ and\ \citenamefont {Chuang}(2010)}]{nielsen_2010}%
  \BibitemOpen
  \bibfield  {author} {\bibinfo {author} {\bibfnamefont {M.~A.}\ \bibnamefont {Nielsen}}\ and\ \bibinfo {author} {\bibfnamefont {I.~L.}\ \bibnamefont {Chuang}},\ }\href {\doibase 10.1017/CBO9780511976667} {\emph {\bibinfo {title} {Quantum Computation and Quantum Information: 10th Anniversary Edition}}}\ (\bibinfo  {publisher} {Cambridge University Press},\ \bibinfo {year} {2010})\BibitemShut {NoStop}%
\bibitem [{\citenamefont {Preskill}(2015)}]{Preskill}%
  \BibitemOpen
  \bibfield  {author} {\bibinfo {author} {\bibfnamefont {J.}~\bibnamefont {Preskill}},\ }in\ \href {\doibase http://theory.caltech.edu/~preskill/ph229/} {\emph {\bibinfo {booktitle} {Lecture Notes for Physics 229:Quantum Information and Computation}}}\ (\bibinfo  {publisher} {CreateSpace Independent Publishing Platform},\ \bibinfo {year} {2015})\BibitemShut {NoStop}%
\bibitem [{\citenamefont {Wilde}(2013)}]{wilde_2013}%
  \BibitemOpen
  \bibfield  {author} {\bibinfo {author} {\bibfnamefont {M.~M.}\ \bibnamefont {Wilde}},\ }\href {\doibase 10.1017/CBO9781139525343} {\emph {\bibinfo {title} {Quantum Information Theory}}}\ (\bibinfo  {publisher} {Cambridge University Press},\ \bibinfo {year} {2013})\BibitemShut {NoStop}%
\bibitem [{\citenamefont {Watrous}(2018)}]{Watrous_2018}%
  \BibitemOpen
  \bibfield  {author} {\bibinfo {author} {\bibfnamefont {J.}~\bibnamefont {Watrous}},\ }\href {\doibase 10.1017/9781316848142} {\emph {\bibinfo {title} {The Theory of Quantum Information}}}\ (\bibinfo  {publisher} {Cambridge University Press},\ \bibinfo {year} {2018})\BibitemShut {NoStop}%
\bibitem [{\citenamefont {Ruskai}\ \emph {et~al.}(2002)\citenamefont {Ruskai}, \citenamefont {Szarek},\ and\ \citenamefont {Werner}}]{Ruskai_LAA_2002}%
  \BibitemOpen
  \bibfield  {author} {\bibinfo {author} {\bibfnamefont {M.~B.}\ \bibnamefont {Ruskai}}, \bibinfo {author} {\bibfnamefont {S.}~\bibnamefont {Szarek}}, \ and\ \bibinfo {author} {\bibfnamefont {E.}~\bibnamefont {Werner}},\ }\href {\doibase 10.1016/S0024-3795(01)00547-X} {\bibfield  {journal} {\bibinfo  {journal} {Linear Algebra and its Applications}\ }\textbf {\bibinfo {volume} {347}},\ \bibinfo {pages} {159} (\bibinfo {year} {2002})}\BibitemShut {NoStop}%
\bibitem [{\citenamefont {Kraus}\ \emph {et~al.}(1983)\citenamefont {Kraus}, \citenamefont {Böhm}, \citenamefont {Dollard},\ and\ \citenamefont {Wootters}}]{krauss_2013}%
  \BibitemOpen
  \bibfield  {author} {\bibinfo {author} {\bibfnamefont {K.}~\bibnamefont {Kraus}}, \bibinfo {author} {\bibfnamefont {A.}~\bibnamefont {Böhm}}, \bibinfo {author} {\bibfnamefont {J.~D.}\ \bibnamefont {Dollard}}, \ and\ \bibinfo {author} {\bibfnamefont {W.~H.}\ \bibnamefont {Wootters}},\ }\href {\doibase https://doi.org/10.1007/3-540-12732-1} {\emph {\bibinfo {title} {States, Effects, and Operations}}}\ (\bibinfo  {publisher} {Springer Berlin, Heidelberg},\ \bibinfo {year} {1983})\BibitemShut {NoStop}%
\bibitem [{\citenamefont {de~Pillis}(1967)}]{dePillis_PJM_1967}%
  \BibitemOpen
  \bibfield  {author} {\bibinfo {author} {\bibfnamefont {J.~E.}\ \bibnamefont {de~Pillis}},\ }\href {\doibase 10.2140/pjm.1967.23.129} {\bibfield  {journal} {\bibinfo  {journal} {Pacific Journal of Mathematics}\ }\textbf {\bibinfo {volume} {23}},\ \bibinfo {pages} {129} (\bibinfo {year} {1967})}\BibitemShut {NoStop}%
\bibitem [{\citenamefont {Jamiołkowski}(1972)}]{Jamiolkowski_RMP_1972}%
  \BibitemOpen
  \bibfield  {author} {\bibinfo {author} {\bibfnamefont {A.}~\bibnamefont {Jamiołkowski}},\ }\href {\doibase https://doi.org/10.1016/0034-4877(72)90011-0} {\bibfield  {journal} {\bibinfo  {journal} {Reports on Mathematical Physics}\ }\textbf {\bibinfo {volume} {3}},\ \bibinfo {pages} {275} (\bibinfo {year} {1972})}\BibitemShut {NoStop}%
\bibitem [{\citenamefont {Choi}(1975)}]{Choi_LAA_1975}%
  \BibitemOpen
  \bibfield  {author} {\bibinfo {author} {\bibfnamefont {M.}~\bibnamefont {Choi}},\ }\href {\doibase https://doi.org/10.1016/0024-3795(75)90075-0} {\bibfield  {journal} {\bibinfo  {journal} {Linear Algebra and its Applications}\ }\textbf {\bibinfo {volume} {10}},\ \bibinfo {pages} {285} (\bibinfo {year} {1975})}\BibitemShut {NoStop}%
\bibitem [{\citenamefont {Paulsen}\ and\ \citenamefont {Shultz}(2013)}]{Paulsen_JMP_2013}%
  \BibitemOpen
  \bibfield  {author} {\bibinfo {author} {\bibfnamefont {V.~I.}\ \bibnamefont {Paulsen}}\ and\ \bibinfo {author} {\bibfnamefont {F.}~\bibnamefont {Shultz}},\ }\href {\doibase 10.1063/1.4812329} {\bibfield  {journal} {\bibinfo  {journal} {Journal of Mathematical Physics}\ }\textbf {\bibinfo {volume} {54}},\ \bibinfo {pages} {072201} (\bibinfo {year} {2013})}\BibitemShut {NoStop}%
\bibitem [{\citenamefont {Kye}(2022)}]{Kye_JMP_2022}%
  \BibitemOpen
  \bibfield  {author} {\bibinfo {author} {\bibfnamefont {S.}~\bibnamefont {Kye}},\ }\href {\doibase 10.1063/5.0107646} {\bibfield  {journal} {\bibinfo  {journal} {Journal of Mathematical Physics}\ }\textbf {\bibinfo {volume} {63}},\ \bibinfo {pages} {092202} (\bibinfo {year} {2022})}\BibitemShut {NoStop}%
\bibitem [{\citenamefont {Holevo}(1973{\natexlab{a}})}]{Holevo_PPI_1973}%
  \BibitemOpen
  \bibfield  {author} {\bibinfo {author} {\bibfnamefont {A.~S.}\ \bibnamefont {Holevo}},\ }\href@noop {} {\bibfield  {journal} {\bibinfo  {journal} {Problemy Pereda\v{c}i Informacii}\ }\textbf {\bibinfo {volume} {9}},\ \bibinfo {pages} {3} (\bibinfo {year} {1973}{\natexlab{a}})}\BibitemShut {NoStop}%
\bibitem [{\citenamefont {Holevo}(1973{\natexlab{b}})}]{Holevo_PPI_1973_2}%
  \BibitemOpen
  \bibfield  {author} {\bibinfo {author} {\bibfnamefont {A.~S.}\ \bibnamefont {Holevo}},\ }\href@noop {} {\bibfield  {journal} {\bibinfo  {journal} {Probl. Peredachi Inf.}\ }\textbf {\bibinfo {volume} {9}},\ \bibinfo {pages} {177} (\bibinfo {year} {1973}{\natexlab{b}})}\BibitemShut {NoStop}%
\bibitem [{\citenamefont {Holevo}(1998)}]{Holevo_IEEE_1998}%
  \BibitemOpen
  \bibfield  {author} {\bibinfo {author} {\bibfnamefont {A.}~\bibnamefont {Holevo}},\ }\href {\doibase 10.1109/18.651037} {\bibfield  {journal} {\bibinfo  {journal} {IEEE Transactions on Information Theory}\ }\textbf {\bibinfo {volume} {44}},\ \bibinfo {pages} {269} (\bibinfo {year} {1998})}\BibitemShut {NoStop}%
\bibitem [{\citenamefont {Schumacher}\ and\ \citenamefont {Westmoreland}(1997)}]{Schumacher_PRA_1997}%
  \BibitemOpen
  \bibfield  {author} {\bibinfo {author} {\bibfnamefont {B.}~\bibnamefont {Schumacher}}\ and\ \bibinfo {author} {\bibfnamefont {M.~D.}\ \bibnamefont {Westmoreland}},\ }\href {\doibase 10.1103/PhysRevA.56.131} {\bibfield  {journal} {\bibinfo  {journal} {Phys. Rev. A}\ }\textbf {\bibinfo {volume} {56}},\ \bibinfo {pages} {131} (\bibinfo {year} {1997})}\BibitemShut {NoStop}%
\bibitem [{\citenamefont {Lloyd}(1997)}]{LLoyd_PRA_1997}%
  \BibitemOpen
  \bibfield  {author} {\bibinfo {author} {\bibfnamefont {S.}~\bibnamefont {Lloyd}},\ }\href {\doibase 10.1103/PhysRevA.55.1613} {\bibfield  {journal} {\bibinfo  {journal} {Phys. Rev. A}\ }\textbf {\bibinfo {volume} {55}},\ \bibinfo {pages} {1613} (\bibinfo {year} {1997})}\BibitemShut {NoStop}%
\bibitem [{\citenamefont {Bennett}\ \emph {et~al.}(1997)\citenamefont {Bennett}, \citenamefont {DiVincenzo},\ and\ \citenamefont {Smolin}}]{Bennett_PRL_1997}%
  \BibitemOpen
  \bibfield  {author} {\bibinfo {author} {\bibfnamefont {C.~H.}\ \bibnamefont {Bennett}}, \bibinfo {author} {\bibfnamefont {D.~P.}\ \bibnamefont {DiVincenzo}}, \ and\ \bibinfo {author} {\bibfnamefont {J.~A.}\ \bibnamefont {Smolin}},\ }\href {\doibase 10.1103/PhysRevLett.78.3217} {\bibfield  {journal} {\bibinfo  {journal} {Physical Review Letters}\ }\textbf {\bibinfo {volume} {78}},\ \bibinfo {pages} {3217} (\bibinfo {year} {1997})}\BibitemShut {NoStop}%
\bibitem [{\citenamefont {DiVincenzo}\ \emph {et~al.}(1998)\citenamefont {DiVincenzo}, \citenamefont {Shor},\ and\ \citenamefont {Smolin}}]{DiVincenzo_PRA_1998}%
  \BibitemOpen
  \bibfield  {author} {\bibinfo {author} {\bibfnamefont {D.~P.}\ \bibnamefont {DiVincenzo}}, \bibinfo {author} {\bibfnamefont {P.~W.}\ \bibnamefont {Shor}}, \ and\ \bibinfo {author} {\bibfnamefont {J.~A.}\ \bibnamefont {Smolin}},\ }\href {\doibase 10.1103/PhysRevA.57.830} {\bibfield  {journal} {\bibinfo  {journal} {Physical Review A}\ }\textbf {\bibinfo {volume} {57}},\ \bibinfo {pages} {830} (\bibinfo {year} {1998})}\BibitemShut {NoStop}%
\bibitem [{\citenamefont {Hastings}(2009)}]{Hastings_Nature_2009}%
  \BibitemOpen
  \bibfield  {author} {\bibinfo {author} {\bibfnamefont {M.~B.}\ \bibnamefont {Hastings}},\ }\href {\doibase 10.1038/nphys1224} {\bibfield  {journal} {\bibinfo  {journal} {Nature Physics}\ }\textbf {\bibinfo {volume} {5}},\ \bibinfo {pages} {255} (\bibinfo {year} {2009})}\BibitemShut {NoStop}%
\bibitem [{\citenamefont {Smith}\ and\ \citenamefont {Yard}(2008)}]{Smith_Science_2008}%
  \BibitemOpen
  \bibfield  {author} {\bibinfo {author} {\bibfnamefont {G.}~\bibnamefont {Smith}}\ and\ \bibinfo {author} {\bibfnamefont {J.}~\bibnamefont {Yard}},\ }\href {\doibase 10.1126/science.1162242} {\bibfield  {journal} {\bibinfo  {journal} {Science}\ }\textbf {\bibinfo {volume} {321}},\ \bibinfo {pages} {1812} (\bibinfo {year} {2008})}\BibitemShut {NoStop}%
\bibitem [{\citenamefont {Cubitt}\ \emph {et~al.}(2011)\citenamefont {Cubitt}, \citenamefont {Chen},\ and\ \citenamefont {Harrow}}]{Cubitt_IEEE_2011}%
  \BibitemOpen
  \bibfield  {author} {\bibinfo {author} {\bibfnamefont {T.~S.}\ \bibnamefont {Cubitt}}, \bibinfo {author} {\bibfnamefont {J.}~\bibnamefont {Chen}}, \ and\ \bibinfo {author} {\bibfnamefont {A.~W.}\ \bibnamefont {Harrow}},\ }\href {\doibase 10.1109/TIT.2011.2169109} {\bibfield  {journal} {\bibinfo  {journal} {IEEE Transactions on Information Theory}\ }\textbf {\bibinfo {volume} {57}},\ \bibinfo {pages} {8114} (\bibinfo {year} {2011})}\BibitemShut {NoStop}%
\bibitem [{\citenamefont {Brandao}\ and\ \citenamefont {Oppenheim}(2013)}]{Brandao_IEEE_2013}%
  \BibitemOpen
  \bibfield  {author} {\bibinfo {author} {\bibfnamefont {F.~G. S.~L.}\ \bibnamefont {Brandao}}\ and\ \bibinfo {author} {\bibfnamefont {J.}~\bibnamefont {Oppenheim}},\ }\href {\doibase 10.1109/TIT.2012.2236911} {\bibfield  {journal} {\bibinfo  {journal} {IEEE Transactions on Information Theory}\ }\textbf {\bibinfo {volume} {59}},\ \bibinfo {pages} {2517} (\bibinfo {year} {2013})}\BibitemShut {NoStop}%
\bibitem [{\citenamefont {Shirokov}\ and\ \citenamefont {Shulman}(2015)}]{Shirokov_CMP_2015}%
  \BibitemOpen
  \bibfield  {author} {\bibinfo {author} {\bibfnamefont {M.~E.}\ \bibnamefont {Shirokov}}\ and\ \bibinfo {author} {\bibfnamefont {T.}~\bibnamefont {Shulman}},\ }\href {\doibase 10.1007/s00220-015-2345-5} {\bibfield  {journal} {\bibinfo  {journal} {Communications in Mathematical Physics}\ }\textbf {\bibinfo {volume} {335}},\ \bibinfo {pages} {1159} (\bibinfo {year} {2015})}\BibitemShut {NoStop}%
\bibitem [{\citenamefont {Bennett}\ \emph {et~al.}(2002)\citenamefont {Bennett}, \citenamefont {Shor}, \citenamefont {Smolin},\ and\ \citenamefont {Thapliyal}}]{Bennett_IEEE_2002}%
  \BibitemOpen
  \bibfield  {author} {\bibinfo {author} {\bibfnamefont {C.~H.}\ \bibnamefont {Bennett}}, \bibinfo {author} {\bibfnamefont {P.~W.}\ \bibnamefont {Shor}}, \bibinfo {author} {\bibfnamefont {J.~A.}\ \bibnamefont {Smolin}}, \ and\ \bibinfo {author} {\bibfnamefont {A.~V.}\ \bibnamefont {Thapliyal}},\ }\href {\doibase 10.1109/TIT.2002.802612} {\bibfield  {journal} {\bibinfo  {journal} {IEEE Transactions on Information Theory}\ }\textbf {\bibinfo {volume} {48}},\ \bibinfo {pages} {2637} (\bibinfo {year} {2002})}\BibitemShut {NoStop}%
\bibitem [{\citenamefont {Devetak}(2005)}]{Devetak_IEEE_2005}%
  \BibitemOpen
  \bibfield  {author} {\bibinfo {author} {\bibfnamefont {I.}~\bibnamefont {Devetak}},\ }\href {\doibase 10.1109/TIT.2004.839515} {\bibfield  {journal} {\bibinfo  {journal} {IEEE Transactions on Information Theory}\ }\textbf {\bibinfo {volume} {51}},\ \bibinfo {pages} {44} (\bibinfo {year} {2005})}\BibitemShut {NoStop}%
\bibitem [{\citenamefont {Shirokov}(2008)}]{Shirokov_TVP_2008}%
  \BibitemOpen
  \bibfield  {author} {\bibinfo {author} {\bibfnamefont {M.~E.}\ \bibnamefont {Shirokov}},\ }\href {\doibase https://doi.org/10.1137/S0040585X97983870} {\bibfield  {journal} {\bibinfo  {journal} {Teor. Veroyatnost. i Primenen.}\ }\textbf {\bibinfo {volume} {53}},\ \bibinfo {pages} {648} (\bibinfo {year} {2008})}\BibitemShut {NoStop}%
\bibitem [{\citenamefont {Davis}\ \emph {et~al.}(2018)\citenamefont {Davis}, \citenamefont {Shirokov},\ and\ \citenamefont {Wilde}}]{Shirokov_PRA_2018}%
  \BibitemOpen
  \bibfield  {author} {\bibinfo {author} {\bibfnamefont {N.}~\bibnamefont {Davis}}, \bibinfo {author} {\bibfnamefont {M.~E.}\ \bibnamefont {Shirokov}}, \ and\ \bibinfo {author} {\bibfnamefont {M.~M.}\ \bibnamefont {Wilde}},\ }\href {\doibase 10.1103/PhysRevA.97.062310} {\bibfield  {journal} {\bibinfo  {journal} {Phys. Rev. A}\ }\textbf {\bibinfo {volume} {97}},\ \bibinfo {pages} {062310} (\bibinfo {year} {2018})}\BibitemShut {NoStop}%
\bibitem [{\citenamefont {Shirokov}(2019)}]{Shirokov_JMP_2019}%
  \BibitemOpen
  \bibfield  {author} {\bibinfo {author} {\bibfnamefont {M.~E.}\ \bibnamefont {Shirokov}},\ }\href {\doibase 10.1088/1751-8121/aaebac} {\bibfield  {journal} {\bibinfo  {journal} {J. Phys. A: Math. Theor.}\ }\textbf {\bibinfo {volume} {52}},\ \bibinfo {pages} {014001} (\bibinfo {year} {2019})}\BibitemShut {NoStop}%
\bibitem [{\citenamefont {Holevo}\ and\ \citenamefont {Shirokov}(2006)}]{Holevo_TPA_2006}%
  \BibitemOpen
  \bibfield  {author} {\bibinfo {author} {\bibfnamefont {A.~S.}\ \bibnamefont {Holevo}}\ and\ \bibinfo {author} {\bibfnamefont {M.~E.}\ \bibnamefont {Shirokov}},\ }\href {\doibase 10.1137/S0040585X97981470} {\bibfield  {journal} {\bibinfo  {journal} {Theory of Probability \& Its Applications}\ }\textbf {\bibinfo {volume} {50}},\ \bibinfo {pages} {86} (\bibinfo {year} {2006})}\BibitemShut {NoStop}%
\bibitem [{\citenamefont {Holevo}(2008)}]{Holevo_PIT_2008}%
  \BibitemOpen
  \bibfield  {author} {\bibinfo {author} {\bibfnamefont {A.~S.}\ \bibnamefont {Holevo}},\ }\href {\doibase 10.1134/S0032946008030010} {\bibfield  {journal} {\bibinfo  {journal} {Problems of Information Transmission}\ }\textbf {\bibinfo {volume} {44}},\ \bibinfo {pages} {171} (\bibinfo {year} {2008})}\BibitemShut {NoStop}%
\bibitem [{\citenamefont {Shirokov}\ and\ \citenamefont {Holevo}(2008)}]{Shirokov_PIT_2008}%
  \BibitemOpen
  \bibfield  {author} {\bibinfo {author} {\bibfnamefont {M.~E.}\ \bibnamefont {Shirokov}}\ and\ \bibinfo {author} {\bibfnamefont {A.~S.}\ \bibnamefont {Holevo}},\ }\href {\doibase 10.1134/S0032946008020014} {\bibfield  {journal} {\bibinfo  {journal} {Problems of Information Transmission}\ }\textbf {\bibinfo {volume} {44}},\ \bibinfo {pages} {73} (\bibinfo {year} {2008})}\BibitemShut {NoStop}%
\bibitem [{\citenamefont {Holevo}(2010)}]{Holevo_DM_2010}%
  \BibitemOpen
  \bibfield  {author} {\bibinfo {author} {\bibfnamefont {A.~S.}\ \bibnamefont {Holevo}},\ }\href {\doibase 10.1134/S1064562410050133} {\bibfield  {journal} {\bibinfo  {journal} {Doklady Mathematics}\ }\textbf {\bibinfo {volume} {82}},\ \bibinfo {pages} {730} (\bibinfo {year} {2010})}\BibitemShut {NoStop}%
\bibitem [{\citenamefont {Landau}\ and\ \citenamefont {Streater}(1993)}]{Landau_LAA_1993}%
  \BibitemOpen
  \bibfield  {author} {\bibinfo {author} {\bibfnamefont {L.}~\bibnamefont {Landau}}\ and\ \bibinfo {author} {\bibfnamefont {R.}~\bibnamefont {Streater}},\ }\href {\doibase 10.1016/0024-3795(93)90274-R} {\bibfield  {journal} {\bibinfo  {journal} {Linear Algebra and its Applications}\ }\textbf {\bibinfo {volume} {193}},\ \bibinfo {pages} {107} (\bibinfo {year} {1993})}\BibitemShut {NoStop}%
\bibitem [{\citenamefont {Mendl}\ and\ \citenamefont {Wolf}(2009)}]{Mendl_CMP_2009}%
  \BibitemOpen
  \bibfield  {author} {\bibinfo {author} {\bibfnamefont {C.~B.}\ \bibnamefont {Mendl}}\ and\ \bibinfo {author} {\bibfnamefont {M.~M.}\ \bibnamefont {Wolf}},\ }\href {\doibase 10.1007/s00220-009-0824-2} {\bibfield  {journal} {\bibinfo  {journal} {Communications in Mathematical Physics}\ }\textbf {\bibinfo {volume} {289}},\ \bibinfo {pages} {1057} (\bibinfo {year} {2009})}\BibitemShut {NoStop}%
\bibitem [{\citenamefont {Haagerup}\ \emph {et~al.}(2021)\citenamefont {Haagerup}, \citenamefont {Musat},\ and\ \citenamefont {Ruskai}}]{Haagerup_AHP_2021}%
  \BibitemOpen
  \bibfield  {author} {\bibinfo {author} {\bibfnamefont {U.}~\bibnamefont {Haagerup}}, \bibinfo {author} {\bibfnamefont {M.}~\bibnamefont {Musat}}, \ and\ \bibinfo {author} {\bibfnamefont {M.~B.}\ \bibnamefont {Ruskai}},\ }\href {\doibase 10.1007/s00023-021-01071-y} {\bibfield  {journal} {\bibinfo  {journal} {Annales Henri Poincaré}\ }\textbf {\bibinfo {volume} {22}},\ \bibinfo {pages} {3455} (\bibinfo {year} {2021})}\BibitemShut {NoStop}%
\bibitem [{\citenamefont {Girard}\ \emph {et~al.}(2022)\citenamefont {Girard}, \citenamefont {Leung}, \citenamefont {Levick}, \citenamefont {Li}, \citenamefont {Paulsen}, \citenamefont {Poon},\ and\ \citenamefont {Watrous}}]{Girard_CMP_2022}%
  \BibitemOpen
  \bibfield  {author} {\bibinfo {author} {\bibfnamefont {M.}~\bibnamefont {Girard}}, \bibinfo {author} {\bibfnamefont {D.}~\bibnamefont {Leung}}, \bibinfo {author} {\bibfnamefont {J.}~\bibnamefont {Levick}}, \bibinfo {author} {\bibfnamefont {C.}~\bibnamefont {Li}}, \bibinfo {author} {\bibfnamefont {V.}~\bibnamefont {Paulsen}}, \bibinfo {author} {\bibfnamefont {Y.~T.}\ \bibnamefont {Poon}}, \ and\ \bibinfo {author} {\bibfnamefont {J.}~\bibnamefont {Watrous}},\ }\href {\doibase 10.1007/s00220-022-04412-y} {\bibfield  {journal} {\bibinfo  {journal} {Communications in Mathematical Physics}\ }\textbf {\bibinfo {volume} {394}},\ \bibinfo {pages} {919} (\bibinfo {year} {2022})}\BibitemShut {NoStop}%
\bibitem [{\citenamefont {Li}\ and\ \citenamefont {Choi}(2023)}]{Li_arXiv_2023}%
  \BibitemOpen
  \bibfield  {author} {\bibinfo {author} {\bibfnamefont {C.-K.}\ \bibnamefont {Li}}\ and\ \bibinfo {author} {\bibfnamefont {M.-D.}\ \bibnamefont {Choi}},\ }\href {https://arxiv.org/abs/2301.01358} {\  (\bibinfo {year} {2023})}\BibitemShut {NoStop}%
\bibitem [{\citenamefont {Bae}\ and\ \citenamefont {Kwek}(2015)}]{Bae_JPA_2015}%
  \BibitemOpen
  \bibfield  {author} {\bibinfo {author} {\bibfnamefont {J.}~\bibnamefont {Bae}}\ and\ \bibinfo {author} {\bibfnamefont {L.-C.}\ \bibnamefont {Kwek}},\ }\href {\doibase 10.1088/1751-8113/48/8/083001} {\bibfield  {journal} {\bibinfo  {journal} {Journal of Physics A: Mathematical and Theoretical}\ }\textbf {\bibinfo {volume} {48}},\ \bibinfo {pages} {083001} (\bibinfo {year} {2015})}\BibitemShut {NoStop}%
\bibitem [{\citenamefont {Kribs}\ \emph {et~al.}(2005)\citenamefont {Kribs}, \citenamefont {Laflamme},\ and\ \citenamefont {Poulin}}]{Kribs_PRL_2005}%
  \BibitemOpen
  \bibfield  {author} {\bibinfo {author} {\bibfnamefont {D.}~\bibnamefont {Kribs}}, \bibinfo {author} {\bibfnamefont {R.}~\bibnamefont {Laflamme}}, \ and\ \bibinfo {author} {\bibfnamefont {D.}~\bibnamefont {Poulin}},\ }\href {\doibase 10.1103/PhysRevLett.94.180501} {\bibfield  {journal} {\bibinfo  {journal} {Phys. Rev. Lett.}\ }\textbf {\bibinfo {volume} {94}},\ \bibinfo {pages} {180501} (\bibinfo {year} {2005})}\BibitemShut {NoStop}%
\bibitem [{\citenamefont {Poulin}(2005)}]{Poulin_PRL_2005}%
  \BibitemOpen
  \bibfield  {author} {\bibinfo {author} {\bibfnamefont {D.}~\bibnamefont {Poulin}},\ }\href {\doibase 10.1103/PhysRevLett.95.230504} {\bibfield  {journal} {\bibinfo  {journal} {Phys. Rev. Lett.}\ }\textbf {\bibinfo {volume} {95}},\ \bibinfo {pages} {230504} (\bibinfo {year} {2005})}\BibitemShut {NoStop}%
\bibitem [{\citenamefont {Bacon}(2006)}]{Bacon_PRA_2006}%
  \BibitemOpen
  \bibfield  {author} {\bibinfo {author} {\bibfnamefont {D.}~\bibnamefont {Bacon}},\ }\href {\doibase 10.1103/PhysRevA.73.012340} {\bibfield  {journal} {\bibinfo  {journal} {Phys. Rev. A}\ }\textbf {\bibinfo {volume} {73}},\ \bibinfo {pages} {012340} (\bibinfo {year} {2006})}\BibitemShut {NoStop}%
\bibitem [{\citenamefont {Nielsen}\ and\ \citenamefont {Poulin}(2007)}]{Nielsen_PRA_2007}%
  \BibitemOpen
  \bibfield  {author} {\bibinfo {author} {\bibfnamefont {M.~A.}\ \bibnamefont {Nielsen}}\ and\ \bibinfo {author} {\bibfnamefont {D.}~\bibnamefont {Poulin}},\ }\href {\doibase 10.1103/PhysRevA.75.064304} {\bibfield  {journal} {\bibinfo  {journal} {Phys. Rev. A}\ }\textbf {\bibinfo {volume} {75}},\ \bibinfo {pages} {064304} (\bibinfo {year} {2007})}\BibitemShut {NoStop}%
\bibitem [{\citenamefont {Memarzadeh}\ and\ \citenamefont {Sanders}(2022)}]{Memarzadeh_PRR_2022}%
  \BibitemOpen
  \bibfield  {author} {\bibinfo {author} {\bibfnamefont {L.}~\bibnamefont {Memarzadeh}}\ and\ \bibinfo {author} {\bibfnamefont {B.~C.}\ \bibnamefont {Sanders}},\ }\href {\doibase 10.1103/PhysRevResearch.4.033206} {\bibfield  {journal} {\bibinfo  {journal} {Physical Review Research}\ }\textbf {\bibinfo {volume} {4}},\ \bibinfo {pages} {033206} (\bibinfo {year} {2022})}\BibitemShut {NoStop}%
\bibitem [{\citenamefont {Mozrzymas}\ \emph {et~al.}(2017)\citenamefont {Mozrzymas}, \citenamefont {Studziński},\ and\ \citenamefont {Datta}}]{Mozrzymas_JMP_2017}%
  \BibitemOpen
  \bibfield  {author} {\bibinfo {author} {\bibfnamefont {M.}~\bibnamefont {Mozrzymas}}, \bibinfo {author} {\bibfnamefont {M.}~\bibnamefont {Studziński}}, \ and\ \bibinfo {author} {\bibfnamefont {N.}~\bibnamefont {Datta}},\ }\href {\doibase 10.1063/1.4983710} {\bibfield  {journal} {\bibinfo  {journal} {Journal of Mathematical Physics}\ }\textbf {\bibinfo {volume} {58}} (\bibinfo {year} {2017}),\ 10.1063/1.4983710}\BibitemShut {NoStop}%
\bibitem [{\citenamefont {Kopszak}\ \emph {et~al.}(2020)\citenamefont {Kopszak}, \citenamefont {Mozrzymas},\ and\ \citenamefont {Studziński}}]{Kopszak_JPA_2020}%
  \BibitemOpen
  \bibfield  {author} {\bibinfo {author} {\bibfnamefont {P.}~\bibnamefont {Kopszak}}, \bibinfo {author} {\bibfnamefont {M.}~\bibnamefont {Mozrzymas}}, \ and\ \bibinfo {author} {\bibfnamefont {M.}~\bibnamefont {Studziński}},\ }\href {\doibase 10.1088/1751-8121/abaa04} {\bibfield  {journal} {\bibinfo  {journal} {Journal of Physics A: Mathematical and Theoretical}\ }\textbf {\bibinfo {volume} {53}},\ \bibinfo {pages} {395306} (\bibinfo {year} {2020})}\BibitemShut {NoStop}%
\bibitem [{\citenamefont {K\"onig}\ and\ \citenamefont {Wehner}(2009)}]{Konig_PRL_2009}%
  \BibitemOpen
  \bibfield  {author} {\bibinfo {author} {\bibfnamefont {R.}~\bibnamefont {K\"onig}}\ and\ \bibinfo {author} {\bibfnamefont {S.}~\bibnamefont {Wehner}},\ }\href {\doibase 10.1103/PhysRevLett.103.070504} {\bibfield  {journal} {\bibinfo  {journal} {Phys. Rev. Lett.}\ }\textbf {\bibinfo {volume} {103}},\ \bibinfo {pages} {070504} (\bibinfo {year} {2009})}\BibitemShut {NoStop}%
\bibitem [{\citenamefont {Datta}\ \emph {et~al.}(2016)\citenamefont {Datta}, \citenamefont {Tomamichel},\ and\ \citenamefont {Wilde}}]{Datta_QIP_2016}%
  \BibitemOpen
  \bibfield  {author} {\bibinfo {author} {\bibfnamefont {N.}~\bibnamefont {Datta}}, \bibinfo {author} {\bibfnamefont {M.}~\bibnamefont {Tomamichel}}, \ and\ \bibinfo {author} {\bibfnamefont {M.~M.}\ \bibnamefont {Wilde}},\ }\href {\doibase 10.1007/s11128-016-1272-5} {\bibfield  {journal} {\bibinfo  {journal} {Quantum Information Processing}\ }\textbf {\bibinfo {volume} {15}},\ \bibinfo {pages} {2569} (\bibinfo {year} {2016})}\BibitemShut {NoStop}%
\bibitem [{\citenamefont {Wilde}\ \emph {et~al.}(2017)\citenamefont {Wilde}, \citenamefont {Tomamichel},\ and\ \citenamefont {Berta}}]{Wilde_IEEE_2017}%
  \BibitemOpen
  \bibfield  {author} {\bibinfo {author} {\bibfnamefont {M.~M.}\ \bibnamefont {Wilde}}, \bibinfo {author} {\bibfnamefont {M.}~\bibnamefont {Tomamichel}}, \ and\ \bibinfo {author} {\bibfnamefont {M.}~\bibnamefont {Berta}},\ }\href {\doibase 10.1109/TIT.2017.2648825} {\bibfield  {journal} {\bibinfo  {journal} {IEEE Transactions on Information Theory}\ }\textbf {\bibinfo {volume} {63}},\ \bibinfo {pages} {1792} (\bibinfo {year} {2017})}\BibitemShut {NoStop}%
\bibitem [{\citenamefont {Liu}\ \emph {et~al.}(2017)\citenamefont {Liu}, \citenamefont {Hu},\ and\ \citenamefont {Lloyd}}]{Liu_PRL_2017}%
  \BibitemOpen
  \bibfield  {author} {\bibinfo {author} {\bibfnamefont {Z.-W.}\ \bibnamefont {Liu}}, \bibinfo {author} {\bibfnamefont {X.}~\bibnamefont {Hu}}, \ and\ \bibinfo {author} {\bibfnamefont {S.}~\bibnamefont {Lloyd}},\ }\href {\doibase 10.1103/PhysRevLett.118.060502} {\bibfield  {journal} {\bibinfo  {journal} {Phys. Rev. Lett.}\ }\textbf {\bibinfo {volume} {118}},\ \bibinfo {pages} {060502} (\bibinfo {year} {2017})}\BibitemShut {NoStop}%
\bibitem [{\citenamefont {Ruskai}(2003)}]{Ruskai_RMP_2003}%
  \BibitemOpen
  \bibfield  {author} {\bibinfo {author} {\bibfnamefont {M.~B.}\ \bibnamefont {Ruskai}},\ }\href {\doibase 10.1142/S0129055X03001710} {\bibfield  {journal} {\bibinfo  {journal} {Reviews in Mathematical Physics}\ }\textbf {\bibinfo {volume} {15}},\ \bibinfo {pages} {643} (\bibinfo {year} {2003})}\BibitemShut {NoStop}%
\bibitem [{\citenamefont {Horodecki}\ \emph {et~al.}(2003)\citenamefont {Horodecki}, \citenamefont {Shor},\ and\ \citenamefont {Ruskai}}]{Horodecki_RMP_2003}%
  \BibitemOpen
  \bibfield  {author} {\bibinfo {author} {\bibfnamefont {M.}~\bibnamefont {Horodecki}}, \bibinfo {author} {\bibfnamefont {P.~W.}\ \bibnamefont {Shor}}, \ and\ \bibinfo {author} {\bibfnamefont {M.~B.}\ \bibnamefont {Ruskai}},\ }\href {\doibase 10.1142/S0129055X03001709} {\bibfield  {journal} {\bibinfo  {journal} {Reviews in Mathematical Physics}\ }\textbf {\bibinfo {volume} {15}},\ \bibinfo {pages} {629} (\bibinfo {year} {2003})}\BibitemShut {NoStop}%
\bibitem [{\citenamefont {Moravčíková}\ and\ \citenamefont {Ziman}(2010{\natexlab{a}})}]{Ziman_JPA_2010}%
  \BibitemOpen
  \bibfield  {author} {\bibinfo {author} {\bibfnamefont {L.}~\bibnamefont {Moravčíková}}\ and\ \bibinfo {author} {\bibfnamefont {M.}~\bibnamefont {Ziman}},\ }\href {\doibase 10.1088/1751-8113/43/27/275306} {\bibfield  {journal} {\bibinfo  {journal} {Journal of Physics A: Mathematical and Theoretical}\ }\textbf {\bibinfo {volume} {43}},\ \bibinfo {pages} {275306} (\bibinfo {year} {2010}{\natexlab{a}})}\BibitemShut {NoStop}%
\bibitem [{\citenamefont {Filippov}\ \emph {et~al.}(2012)\citenamefont {Filippov}, \citenamefont {Ryb\'ar},\ and\ \citenamefont {Ziman}}]{Filipov_PRA_2012}%
  \BibitemOpen
  \bibfield  {author} {\bibinfo {author} {\bibfnamefont {S.~N.}\ \bibnamefont {Filippov}}, \bibinfo {author} {\bibfnamefont {T.}~\bibnamefont {Ryb\'ar}}, \ and\ \bibinfo {author} {\bibfnamefont {M.}~\bibnamefont {Ziman}},\ }\href {\doibase 10.1103/PhysRevA.85.012303} {\bibfield  {journal} {\bibinfo  {journal} {Phys. Rev. A}\ }\textbf {\bibinfo {volume} {85}},\ \bibinfo {pages} {012303} (\bibinfo {year} {2012})}\BibitemShut {NoStop}%
\bibitem [{\citenamefont {Filippov}\ \emph {et~al.}(2013)\citenamefont {Filippov}, \citenamefont {Melnikov},\ and\ \citenamefont {Ziman}}]{Filipov_PRA_2013}%
  \BibitemOpen
  \bibfield  {author} {\bibinfo {author} {\bibfnamefont {S.~N.}\ \bibnamefont {Filippov}}, \bibinfo {author} {\bibfnamefont {A.~A.}\ \bibnamefont {Melnikov}}, \ and\ \bibinfo {author} {\bibfnamefont {M.}~\bibnamefont {Ziman}},\ }\href {\doibase 10.1103/PhysRevA.88.062328} {\bibfield  {journal} {\bibinfo  {journal} {Phys. Rev. A}\ }\textbf {\bibinfo {volume} {88}},\ \bibinfo {pages} {062328} (\bibinfo {year} {2013})}\BibitemShut {NoStop}%
\bibitem [{\citenamefont {Pal}\ and\ \citenamefont {Ghosh}(2015)}]{Pal_JPA_2015}%
  \BibitemOpen
  \bibfield  {author} {\bibinfo {author} {\bibfnamefont {R.}~\bibnamefont {Pal}}\ and\ \bibinfo {author} {\bibfnamefont {S.}~\bibnamefont {Ghosh}},\ }\href {\doibase 10.1088/1751-8113/48/15/155302} {\bibfield  {journal} {\bibinfo  {journal} {Journal of Physics A: Mathematical and Theoretical}\ }\textbf {\bibinfo {volume} {48}},\ \bibinfo {pages} {155302} (\bibinfo {year} {2015})}\BibitemShut {NoStop}%
\bibitem [{\citenamefont {Luo}\ \emph {et~al.}(2022)\citenamefont {Luo}, \citenamefont {Li},\ and\ \citenamefont {Xi}}]{Luo_QIP_2022}%
  \BibitemOpen
  \bibfield  {author} {\bibinfo {author} {\bibfnamefont {Y.}~\bibnamefont {Luo}}, \bibinfo {author} {\bibfnamefont {Y.}~\bibnamefont {Li}}, \ and\ \bibinfo {author} {\bibfnamefont {Z.}~\bibnamefont {Xi}},\ }\href {\doibase 10.1007/s11128-022-03511-y} {\bibfield  {journal} {\bibinfo  {journal} {Quantum Information Processing}\ }\textbf {\bibinfo {volume} {21}},\ \bibinfo {pages} {176} (\bibinfo {year} {2022})}\BibitemShut {NoStop}%
\bibitem [{\citenamefont {Muhuri}\ \emph {et~al.}(2023)\citenamefont {Muhuri}, \citenamefont {Patra}, \citenamefont {Gupta},\ and\ \citenamefont {De}}]{Muhuri_arXiv_2023}%
  \BibitemOpen
  \bibfield  {author} {\bibinfo {author} {\bibfnamefont {A.}~\bibnamefont {Muhuri}}, \bibinfo {author} {\bibfnamefont {A.}~\bibnamefont {Patra}}, \bibinfo {author} {\bibfnamefont {R.}~\bibnamefont {Gupta}}, \ and\ \bibinfo {author} {\bibfnamefont {A.~S.}\ \bibnamefont {De}},\ }\href {https://doi.org/10.48550/arXiv.2309.03108} {\bibfield  {journal} {\bibinfo  {journal} {arXiv: 2309.03108}\ } (\bibinfo {year} {2023})}\BibitemShut {NoStop}%
\bibitem [{\citenamefont {Bennett}\ and\ \citenamefont {Wiesner}(1992)}]{Bennett_PRL_1992}%
  \BibitemOpen
  \bibfield  {author} {\bibinfo {author} {\bibfnamefont {C.~H.}\ \bibnamefont {Bennett}}\ and\ \bibinfo {author} {\bibfnamefont {S.~J.}\ \bibnamefont {Wiesner}},\ }\href {\doibase 10.1103/PhysRevLett.69.2881} {\bibfield  {journal} {\bibinfo  {journal} {Physical Review Letters}\ }\textbf {\bibinfo {volume} {69}},\ \bibinfo {pages} {2881} (\bibinfo {year} {1992})}\BibitemShut {NoStop}%
\bibitem [{\citenamefont {Bennett}\ \emph {et~al.}(1993)\citenamefont {Bennett}, \citenamefont {Brassard}, \citenamefont {Crépeau}, \citenamefont {Jozsa}, \citenamefont {Peres},\ and\ \citenamefont {Wootters}}]{Bennett_PRL_1993}%
  \BibitemOpen
  \bibfield  {author} {\bibinfo {author} {\bibfnamefont {C.~H.}\ \bibnamefont {Bennett}}, \bibinfo {author} {\bibfnamefont {G.}~\bibnamefont {Brassard}}, \bibinfo {author} {\bibfnamefont {C.}~\bibnamefont {Crépeau}}, \bibinfo {author} {\bibfnamefont {R.}~\bibnamefont {Jozsa}}, \bibinfo {author} {\bibfnamefont {A.}~\bibnamefont {Peres}}, \ and\ \bibinfo {author} {\bibfnamefont {W.~K.}\ \bibnamefont {Wootters}},\ }\href {\doibase 10.1103/PhysRevLett.70.1895} {\bibfield  {journal} {\bibinfo  {journal} {Physical Review Letters}\ }\textbf {\bibinfo {volume} {70}},\ \bibinfo {pages} {1895} (\bibinfo {year} {1993})}\BibitemShut {NoStop}%
\bibitem [{\citenamefont {Reck}\ \emph {et~al.}(1994)\citenamefont {Reck}, \citenamefont {Zeilinger}, \citenamefont {Bernstein},\ and\ \citenamefont {Bertani}}]{Reck_PRL_1994}%
  \BibitemOpen
  \bibfield  {author} {\bibinfo {author} {\bibfnamefont {M.}~\bibnamefont {Reck}}, \bibinfo {author} {\bibfnamefont {A.}~\bibnamefont {Zeilinger}}, \bibinfo {author} {\bibfnamefont {H.~J.}\ \bibnamefont {Bernstein}}, \ and\ \bibinfo {author} {\bibfnamefont {P.}~\bibnamefont {Bertani}},\ }\href {\doibase 10.1103/PhysRevLett.73.58} {\bibfield  {journal} {\bibinfo  {journal} {Phys. Rev. Lett.}\ }\textbf {\bibinfo {volume} {73}},\ \bibinfo {pages} {58} (\bibinfo {year} {1994})}\BibitemShut {NoStop}%
\bibitem [{\citenamefont {Deutsch}\ \emph {et~al.}(1995)\citenamefont {Deutsch}, \citenamefont {Barenco},\ and\ \citenamefont {Ekert}}]{Deutch_PRSL_1995}%
  \BibitemOpen
  \bibfield  {author} {\bibinfo {author} {\bibfnamefont {D.~E.}\ \bibnamefont {Deutsch}}, \bibinfo {author} {\bibfnamefont {A.}~\bibnamefont {Barenco}}, \ and\ \bibinfo {author} {\bibfnamefont {A.}~\bibnamefont {Ekert}},\ }\href {\doibase 10.1098/rspa.1995.0065} {\bibfield  {journal} {\bibinfo  {journal} {Proceedings of the Royal Society of London. Series A: Mathematical and Physical Sciences}\ }\textbf {\bibinfo {volume} {449}},\ \bibinfo {pages} {669} (\bibinfo {year} {1995})}\BibitemShut {NoStop}%
\bibitem [{\citenamefont {DiVincenzo}(1995)}]{DiVincenzo_PRA_1995}%
  \BibitemOpen
  \bibfield  {author} {\bibinfo {author} {\bibfnamefont {D.~P.}\ \bibnamefont {DiVincenzo}},\ }\href {\doibase 10.1103/PhysRevA.51.1015} {\bibfield  {journal} {\bibinfo  {journal} {Phys. Rev. A}\ }\textbf {\bibinfo {volume} {51}},\ \bibinfo {pages} {1015} (\bibinfo {year} {1995})}\BibitemShut {NoStop}%
\bibitem [{\citenamefont {Lloyd}(1995)}]{Lloyd_PRL_1995}%
  \BibitemOpen
  \bibfield  {author} {\bibinfo {author} {\bibfnamefont {S.}~\bibnamefont {Lloyd}},\ }\href {\doibase 10.1103/PhysRevLett.75.346} {\bibfield  {journal} {\bibinfo  {journal} {Phys. Rev. Lett.}\ }\textbf {\bibinfo {volume} {75}},\ \bibinfo {pages} {346} (\bibinfo {year} {1995})}\BibitemShut {NoStop}%
\bibitem [{\citenamefont {Boykin}\ \emph {et~al.}(1999)\citenamefont {Boykin}, \citenamefont {Mor}, \citenamefont {Pulver}, \citenamefont {Roychowdhury},\ and\ \citenamefont {Vatan}}]{Boykin_IEEECS_1999}%
  \BibitemOpen
  \bibfield  {author} {\bibinfo {author} {\bibfnamefont {P.}~\bibnamefont {Boykin}}, \bibinfo {author} {\bibfnamefont {T.}~\bibnamefont {Mor}}, \bibinfo {author} {\bibfnamefont {M.}~\bibnamefont {Pulver}}, \bibinfo {author} {\bibfnamefont {V.}~\bibnamefont {Roychowdhury}}, \ and\ \bibinfo {author} {\bibfnamefont {F.}~\bibnamefont {Vatan}},\ }in\ \href {\doibase 10.1109/SFFCS.1999.814621} {\emph {\bibinfo {booktitle} {40th Annual Symposium on Foundations of Computer Science (Cat. No.99CB37039)}}}\ (\bibinfo  {publisher} {IEEE Comput. Soc},\ \bibinfo {year} {1999})\ pp.\ \bibinfo {pages} {486--494}\BibitemShut {NoStop}%
\bibitem [{\citenamefont {Grover}(1996)}]{Grover_ACM_1996}%
  \BibitemOpen
  \bibfield  {author} {\bibinfo {author} {\bibfnamefont {L.~K.}\ \bibnamefont {Grover}},\ }in\ \href {\doibase 10.1145/237814.237866} {\emph {\bibinfo {booktitle} {Proceedings of the twenty-eighth annual ACM symposium on Theory of computing - STOC '96}}}\ (\bibinfo  {publisher} {ACM Press},\ \bibinfo {year} {1996})\ pp.\ \bibinfo {pages} {212--219}\BibitemShut {NoStop}%
\bibitem [{\citenamefont {Shor}(1997)}]{Shor_SIAM_1997}%
  \BibitemOpen
  \bibfield  {author} {\bibinfo {author} {\bibfnamefont {P.~W.}\ \bibnamefont {Shor}},\ }\href {\doibase 10.1137/S0097539795293172} {\bibfield  {journal} {\bibinfo  {journal} {SIAM Journal on Computing}\ }\textbf {\bibinfo {volume} {26}},\ \bibinfo {pages} {1484} (\bibinfo {year} {1997})}\BibitemShut {NoStop}%
\bibitem [{\citenamefont {Jozsa}(1998)}]{Jozsa_PRSL_1998}%
  \BibitemOpen
  \bibfield  {author} {\bibinfo {author} {\bibfnamefont {R.}~\bibnamefont {Jozsa}},\ }\href {\doibase 10.1098/rspa.1998.0163} {\bibfield  {journal} {\bibinfo  {journal} {Proceedings of the Royal Society of London. Series A: Mathematical, Physical and Engineering Sciences}\ }\textbf {\bibinfo {volume} {454}},\ \bibinfo {pages} {323} (\bibinfo {year} {1998})}\BibitemShut {NoStop}%
\bibitem [{\citenamefont {Shor}(1996)}]{Shor_IEEE_1996}%
  \BibitemOpen
  \bibfield  {author} {\bibinfo {author} {\bibfnamefont {P.}~\bibnamefont {Shor}}\ }(\bibinfo  {publisher} {IEEE Comput. Soc. Press},\ \bibinfo {year} {1996})\ pp.\ \bibinfo {pages} {56--65}\BibitemShut {NoStop}%
\bibitem [{\citenamefont {Gottesman}(1996)}]{Gottesman_PRA_1996}%
  \BibitemOpen
  \bibfield  {author} {\bibinfo {author} {\bibfnamefont {D.}~\bibnamefont {Gottesman}},\ }\href {\doibase 10.1103/PhysRevA.54.1862} {\bibfield  {journal} {\bibinfo  {journal} {Phys. Rev. A}\ }\textbf {\bibinfo {volume} {54}},\ \bibinfo {pages} {1862} (\bibinfo {year} {1996})}\BibitemShut {NoStop}%
\bibitem [{\citenamefont {DiVincenzo}\ and\ \citenamefont {Shor}(1996)}]{DiVincenzo_PRL_1996}%
  \BibitemOpen
  \bibfield  {author} {\bibinfo {author} {\bibfnamefont {D.~P.}\ \bibnamefont {DiVincenzo}}\ and\ \bibinfo {author} {\bibfnamefont {P.~W.}\ \bibnamefont {Shor}},\ }\href {\doibase 10.1103/PhysRevLett.77.3260} {\bibfield  {journal} {\bibinfo  {journal} {Phys. Rev. Lett.}\ }\textbf {\bibinfo {volume} {77}},\ \bibinfo {pages} {3260} (\bibinfo {year} {1996})}\BibitemShut {NoStop}%
\bibitem [{\citenamefont {Gottesman}(1998)}]{Gottessman_PRA_1998}%
  \BibitemOpen
  \bibfield  {author} {\bibinfo {author} {\bibfnamefont {D.}~\bibnamefont {Gottesman}},\ }\href {\doibase 10.1103/PhysRevA.57.127} {\bibfield  {journal} {\bibinfo  {journal} {Phys. Rev. A}\ }\textbf {\bibinfo {volume} {57}},\ \bibinfo {pages} {127} (\bibinfo {year} {1998})}\BibitemShut {NoStop}%
\bibitem [{\citenamefont {Eastin}\ and\ \citenamefont {Knill}(2009)}]{Eastin_PRL_2009}%
  \BibitemOpen
  \bibfield  {author} {\bibinfo {author} {\bibfnamefont {B.}~\bibnamefont {Eastin}}\ and\ \bibinfo {author} {\bibfnamefont {E.}~\bibnamefont {Knill}},\ }\href {\doibase 10.1103/PhysRevLett.102.110502} {\bibfield  {journal} {\bibinfo  {journal} {Phys. Rev. Lett.}\ }\textbf {\bibinfo {volume} {102}},\ \bibinfo {pages} {110502} (\bibinfo {year} {2009})}\BibitemShut {NoStop}%
\bibitem [{\citenamefont {Gottesman}(1999)}]{Gottesman_1999}%
  \BibitemOpen
  \bibfield  {author} {\bibinfo {author} {\bibfnamefont {D.}~\bibnamefont {Gottesman}},\ }in\ \href {\doibase https://doi.org/10.48550/arXiv.quant-ph/9807006} {\emph {\bibinfo {booktitle} {Group22: Proceedings of the XXII International Colloquium on Group Theoretical Methods in Physics}}},\ \bibinfo {editor} {edited by\ \bibinfo {editor} {\bibfnamefont {S.~P.}\ \bibnamefont {Corney}}, \bibinfo {editor} {\bibfnamefont {R.}~\bibnamefont {Delbourgo}}, \ and\ \bibinfo {editor} {\bibfnamefont {P.~D.}\ \bibnamefont {Jarvis}}}\ (\bibinfo  {publisher} {Cambridge, MA, International Press},\ \bibinfo {year} {1999})\ pp.\ \bibinfo {pages} {32--43}\BibitemShut {NoStop}%
\bibitem [{\citenamefont {Gottesman}\ and\ \citenamefont {Chuang}(1999)}]{Gottesman_Nature_1999}%
  \BibitemOpen
  \bibfield  {author} {\bibinfo {author} {\bibfnamefont {D.}~\bibnamefont {Gottesman}}\ and\ \bibinfo {author} {\bibfnamefont {I.~L.}\ \bibnamefont {Chuang}},\ }\href {\doibase 10.1038/46503} {\bibfield  {journal} {\bibinfo  {journal} {Nature}\ }\textbf {\bibinfo {volume} {402}},\ \bibinfo {pages} {390} (\bibinfo {year} {1999})}\BibitemShut {NoStop}%
\bibitem [{\citenamefont {Mari}\ and\ \citenamefont {Eisert}(2012)}]{mari2012positive}%
  \BibitemOpen
  \bibfield  {author} {\bibinfo {author} {\bibfnamefont {A.}~\bibnamefont {Mari}}\ and\ \bibinfo {author} {\bibfnamefont {J.}~\bibnamefont {Eisert}},\ }\href {\doibase 10.1103/PhysRevLett.109.230503} {\bibfield  {journal} {\bibinfo  {journal} {Physical review letters}\ }\textbf {\bibinfo {volume} {109}},\ \bibinfo {pages} {230503} (\bibinfo {year} {2012})}\BibitemShut {NoStop}%
\bibitem [{\citenamefont {Veitch}\ \emph {et~al.}(2012)\citenamefont {Veitch}, \citenamefont {Ferrie}, \citenamefont {Gross},\ and\ \citenamefont {Emerson}}]{veitch2012negative}%
  \BibitemOpen
  \bibfield  {author} {\bibinfo {author} {\bibfnamefont {V.}~\bibnamefont {Veitch}}, \bibinfo {author} {\bibfnamefont {C.}~\bibnamefont {Ferrie}}, \bibinfo {author} {\bibfnamefont {D.}~\bibnamefont {Gross}}, \ and\ \bibinfo {author} {\bibfnamefont {J.}~\bibnamefont {Emerson}},\ }\href {\doibase 10.1088/1367-2630/14/11/113011} {\bibfield  {journal} {\bibinfo  {journal} {New Journal of Physics}\ }\textbf {\bibinfo {volume} {14}},\ \bibinfo {pages} {113011} (\bibinfo {year} {2012})}\BibitemShut {NoStop}%
\bibitem [{\citenamefont {Veitch}\ \emph {et~al.}(2014)\citenamefont {Veitch}, \citenamefont {Mousavian}, \citenamefont {Gottesman},\ and\ \citenamefont {Emerson}}]{Veitch_NJP_2014}%
  \BibitemOpen
  \bibfield  {author} {\bibinfo {author} {\bibfnamefont {V.}~\bibnamefont {Veitch}}, \bibinfo {author} {\bibfnamefont {S.~A.~H.}\ \bibnamefont {Mousavian}}, \bibinfo {author} {\bibfnamefont {D.}~\bibnamefont {Gottesman}}, \ and\ \bibinfo {author} {\bibfnamefont {J.}~\bibnamefont {Emerson}},\ }\href {\doibase 10.1088/1367-2630/16/1/013009} {\bibfield  {journal} {\bibinfo  {journal} {New Journal of Physics}\ }\textbf {\bibinfo {volume} {16}},\ \bibinfo {pages} {013009} (\bibinfo {year} {2014})}\BibitemShut {NoStop}%
\bibitem [{\citenamefont {Kraus}\ and\ \citenamefont {Cirac}(2001)}]{Kraus_PRA_2001}%
  \BibitemOpen
  \bibfield  {author} {\bibinfo {author} {\bibfnamefont {B.}~\bibnamefont {Kraus}}\ and\ \bibinfo {author} {\bibfnamefont {J.~I.}\ \bibnamefont {Cirac}},\ }\href {\doibase 10.1103/PhysRevA.63.062309} {\bibfield  {journal} {\bibinfo  {journal} {Phys. Rev. A}\ }\textbf {\bibinfo {volume} {63}},\ \bibinfo {pages} {062309} (\bibinfo {year} {2001})}\BibitemShut {NoStop}%
\bibitem [{\citenamefont {Leifer}\ \emph {et~al.}(2003)\citenamefont {Leifer}, \citenamefont {Henderson},\ and\ \citenamefont {Linden}}]{Leifer_PRA_2003}%
  \BibitemOpen
  \bibfield  {author} {\bibinfo {author} {\bibfnamefont {M.~S.}\ \bibnamefont {Leifer}}, \bibinfo {author} {\bibfnamefont {L.}~\bibnamefont {Henderson}}, \ and\ \bibinfo {author} {\bibfnamefont {N.}~\bibnamefont {Linden}},\ }\href {\doibase 10.1103/PhysRevA.67.012306} {\bibfield  {journal} {\bibinfo  {journal} {Phys. Rev. A}\ }\textbf {\bibinfo {volume} {67}},\ \bibinfo {pages} {012306} (\bibinfo {year} {2003})}\BibitemShut {NoStop}%
\bibitem [{\citenamefont {Chefles}(2005)}]{Chefles_PRA_2005}%
  \BibitemOpen
  \bibfield  {author} {\bibinfo {author} {\bibfnamefont {A.}~\bibnamefont {Chefles}},\ }\href {\doibase 10.1103/PhysRevA.72.042332} {\bibfield  {journal} {\bibinfo  {journal} {Phys. Rev. A}\ }\textbf {\bibinfo {volume} {72}},\ \bibinfo {pages} {042332} (\bibinfo {year} {2005})}\BibitemShut {NoStop}%
\bibitem [{\citenamefont {Ku\ifmmode~\acute{s}\else \'{s}\fi{}}\ and\ \citenamefont {\ifmmode~\dot{Z}\else \.{Z}\fi{}yczkowski}(2001)}]{Kush_PRA_2001}%
  \BibitemOpen
  \bibfield  {author} {\bibinfo {author} {\bibfnamefont {M.}~\bibnamefont {Ku\ifmmode~\acute{s}\else \'{s}\fi{}}}\ and\ \bibinfo {author} {\bibfnamefont {K.}~\bibnamefont {\ifmmode~\dot{Z}\else \.{Z}\fi{}yczkowski}},\ }\href {\doibase 10.1103/PhysRevA.63.032307} {\bibfield  {journal} {\bibinfo  {journal} {Phys. Rev. A}\ }\textbf {\bibinfo {volume} {63}},\ \bibinfo {pages} {032307} (\bibinfo {year} {2001})}\BibitemShut {NoStop}%
\bibitem [{\citenamefont {Verstraete}\ \emph {et~al.}(2001)\citenamefont {Verstraete}, \citenamefont {Audenaert},\ and\ \citenamefont {De~Moor}}]{Verstraete_PRA_2001}%
  \BibitemOpen
  \bibfield  {author} {\bibinfo {author} {\bibfnamefont {F.}~\bibnamefont {Verstraete}}, \bibinfo {author} {\bibfnamefont {K.}~\bibnamefont {Audenaert}}, \ and\ \bibinfo {author} {\bibfnamefont {B.}~\bibnamefont {De~Moor}},\ }\href {\doibase 10.1103/PhysRevA.64.012316} {\bibfield  {journal} {\bibinfo  {journal} {Phys. Rev. A}\ }\textbf {\bibinfo {volume} {64}},\ \bibinfo {pages} {012316} (\bibinfo {year} {2001})}\BibitemShut {NoStop}%
\bibitem [{\citenamefont {Horodecki}\ \emph {et~al.}(2009)\citenamefont {Horodecki}, \citenamefont {Horodecki}, \citenamefont {Horodecki},\ and\ \citenamefont {Horodecki}}]{Horodecki_RMP_2009}%
  \BibitemOpen
  \bibfield  {author} {\bibinfo {author} {\bibfnamefont {R.}~\bibnamefont {Horodecki}}, \bibinfo {author} {\bibfnamefont {P.}~\bibnamefont {Horodecki}}, \bibinfo {author} {\bibfnamefont {M.}~\bibnamefont {Horodecki}}, \ and\ \bibinfo {author} {\bibfnamefont {K.}~\bibnamefont {Horodecki}},\ }\href {\doibase 10.1103/RevModPhys.81.865} {\bibfield  {journal} {\bibinfo  {journal} {Rev. Mod. Phys.}\ }\textbf {\bibinfo {volume} {81}},\ \bibinfo {pages} {865} (\bibinfo {year} {2009})}\BibitemShut {NoStop}%
\bibitem [{\citenamefont {Kuroiwa}\ \emph {et~al.}(2024)\citenamefont {Kuroiwa}, \citenamefont {Takagi}, \citenamefont {Adesso},\ and\ \citenamefont {Yamasaki}}]{Kuroiwa_PRL_2024}%
  \BibitemOpen
  \bibfield  {author} {\bibinfo {author} {\bibfnamefont {K.}~\bibnamefont {Kuroiwa}}, \bibinfo {author} {\bibfnamefont {R.}~\bibnamefont {Takagi}}, \bibinfo {author} {\bibfnamefont {G.}~\bibnamefont {Adesso}}, \ and\ \bibinfo {author} {\bibfnamefont {H.}~\bibnamefont {Yamasaki}},\ }\href {\doibase 10.1103/PhysRevLett.132.150201} {\bibfield  {journal} {\bibinfo  {journal} {Phys. Rev. Lett.}\ }\textbf {\bibinfo {volume} {132}},\ \bibinfo {pages} {150201} (\bibinfo {year} {2024})}\BibitemShut {NoStop}%
\bibitem [{\citenamefont {Salazar}\ \emph {et~al.}(2024)\citenamefont {Salazar}, \citenamefont {Czartowski}, \citenamefont {Rodríguez}, \citenamefont {Rajchel-Mieldzioć}, \citenamefont {Horodecki},\ and\ \citenamefont {Życzkowski}}]{Salazar_arXiv_2024}%
  \BibitemOpen
  \bibfield  {author} {\bibinfo {author} {\bibfnamefont {R.}~\bibnamefont {Salazar}}, \bibinfo {author} {\bibfnamefont {J.}~\bibnamefont {Czartowski}}, \bibinfo {author} {\bibfnamefont {R.~R.}\ \bibnamefont {Rodríguez}}, \bibinfo {author} {\bibfnamefont {G.}~\bibnamefont {Rajchel-Mieldzioć}}, \bibinfo {author} {\bibfnamefont {P.}~\bibnamefont {Horodecki}}, \ and\ \bibinfo {author} {\bibfnamefont {K.}~\bibnamefont {Życzkowski}},\ }\href {https://doi.org/10.48550/arXiv.2405.05785} {\bibfield  {journal} {\bibinfo  {journal} {arXiv: 2405.05785}\ } (\bibinfo {year} {2024})}\BibitemShut {NoStop}%
\bibitem [{\citenamefont {Gour}(2024)}]{Gour_arXiv_2024}%
  \BibitemOpen
  \bibfield  {author} {\bibinfo {author} {\bibfnamefont {G.}~\bibnamefont {Gour}},\ }\href {https://doi.org/10.48550/arXiv.2402.05474} {\bibfield  {journal} {\bibinfo  {journal} {arXiv: 2402.05474}\ } (\bibinfo {year} {2024})}\BibitemShut {NoStop}%
\bibitem [{\citenamefont {Bravyi}\ and\ \citenamefont {Kitaev}(2005)}]{Bravyi_PRA_2005}%
  \BibitemOpen
  \bibfield  {author} {\bibinfo {author} {\bibfnamefont {S.}~\bibnamefont {Bravyi}}\ and\ \bibinfo {author} {\bibfnamefont {A.}~\bibnamefont {Kitaev}},\ }\href {\doibase 10.1103/PhysRevA.71.022316} {\bibfield  {journal} {\bibinfo  {journal} {Phys. Rev. A}\ }\textbf {\bibinfo {volume} {71}},\ \bibinfo {pages} {022316} (\bibinfo {year} {2005})}\BibitemShut {NoStop}%
\bibitem [{\citenamefont {Howard}\ and\ \citenamefont {Campbell}(2017)}]{Howard_PRL_2017}%
  \BibitemOpen
  \bibfield  {author} {\bibinfo {author} {\bibfnamefont {M.}~\bibnamefont {Howard}}\ and\ \bibinfo {author} {\bibfnamefont {E.}~\bibnamefont {Campbell}},\ }\href {\doibase 10.1103/PhysRevLett.118.090501} {\bibfield  {journal} {\bibinfo  {journal} {Phys. Rev. Lett.}\ }\textbf {\bibinfo {volume} {118}},\ \bibinfo {pages} {090501} (\bibinfo {year} {2017})}\BibitemShut {NoStop}%
\bibitem [{\citenamefont {Calcluth}\ \emph {et~al.}(2024)\citenamefont {Calcluth}, \citenamefont {Reichel}, \citenamefont {Ferraro},\ and\ \citenamefont {Ferrini}}]{Calcluth_PRX_2024}%
  \BibitemOpen
  \bibfield  {author} {\bibinfo {author} {\bibfnamefont {C.}~\bibnamefont {Calcluth}}, \bibinfo {author} {\bibfnamefont {N.}~\bibnamefont {Reichel}}, \bibinfo {author} {\bibfnamefont {A.}~\bibnamefont {Ferraro}}, \ and\ \bibinfo {author} {\bibfnamefont {G.}~\bibnamefont {Ferrini}},\ }\href {\doibase 10.1103/PRXQuantum.5.020337} {\bibfield  {journal} {\bibinfo  {journal} {PRX Quantum}\ }\textbf {\bibinfo {volume} {5}},\ \bibinfo {pages} {020337} (\bibinfo {year} {2024})}\BibitemShut {NoStop}%
\bibitem [{\citenamefont {Grimsmo}\ \emph {et~al.}(2020)\citenamefont {Grimsmo}, \citenamefont {Combes},\ and\ \citenamefont {Baragiola}}]{Grismo_PRX_2020}%
  \BibitemOpen
  \bibfield  {author} {\bibinfo {author} {\bibfnamefont {A.~L.}\ \bibnamefont {Grimsmo}}, \bibinfo {author} {\bibfnamefont {J.}~\bibnamefont {Combes}}, \ and\ \bibinfo {author} {\bibfnamefont {B.~Q.}\ \bibnamefont {Baragiola}},\ }\href {\doibase 10.1103/PhysRevX.10.011058} {\bibfield  {journal} {\bibinfo  {journal} {Phys. Rev. X}\ }\textbf {\bibinfo {volume} {10}},\ \bibinfo {pages} {011058} (\bibinfo {year} {2020})}\BibitemShut {NoStop}%
\bibitem [{\citenamefont {Reichardt}(2009)}]{Reichardt_QIC_2009}%
  \BibitemOpen
  \bibfield  {author} {\bibinfo {author} {\bibfnamefont {B.}~\bibnamefont {Reichardt}},\ }\href {\doibase 10.26421/QIC9.11-12-7} {\bibfield  {journal} {\bibinfo  {journal} {Quantum Information and Computation}\ }\textbf {\bibinfo {volume} {9}},\ \bibinfo {pages} {1030} (\bibinfo {year} {2009})}\BibitemShut {NoStop}%
\bibitem [{\citenamefont {Seddon}\ \emph {et~al.}(2021)\citenamefont {Seddon}, \citenamefont {Regula}, \citenamefont {Pashayan}, \citenamefont {Ouyang},\ and\ \citenamefont {Campbell}}]{Seddon_PRX_2021}%
  \BibitemOpen
  \bibfield  {author} {\bibinfo {author} {\bibfnamefont {J.~R.}\ \bibnamefont {Seddon}}, \bibinfo {author} {\bibfnamefont {B.}~\bibnamefont {Regula}}, \bibinfo {author} {\bibfnamefont {H.}~\bibnamefont {Pashayan}}, \bibinfo {author} {\bibfnamefont {Y.}~\bibnamefont {Ouyang}}, \ and\ \bibinfo {author} {\bibfnamefont {E.~T.}\ \bibnamefont {Campbell}},\ }\href {\doibase 10.1103/PRXQuantum.2.010345} {\bibfield  {journal} {\bibinfo  {journal} {PRX Quantum}\ }\textbf {\bibinfo {volume} {2}},\ \bibinfo {pages} {010345} (\bibinfo {year} {2021})}\BibitemShut {NoStop}%
\bibitem [{\citenamefont {Leung}\ and\ \citenamefont {Watrous}(2017)}]{Leung_Quantum_2017}%
  \BibitemOpen
  \bibfield  {author} {\bibinfo {author} {\bibfnamefont {D.}~\bibnamefont {Leung}}\ and\ \bibinfo {author} {\bibfnamefont {J.}~\bibnamefont {Watrous}},\ }\href {\doibase 10.22331/q-2017-09-19-28} {\bibfield  {journal} {\bibinfo  {journal} {Quantum}\ }\textbf {\bibinfo {volume} {1}},\ \bibinfo {pages} {28} (\bibinfo {year} {2017})}\BibitemShut {NoStop}%
\bibitem [{\citenamefont {Filippov}\ and\ \citenamefont {Ziman}(2013)}]{Filippov_PRA_2013_2}%
  \BibitemOpen
  \bibfield  {author} {\bibinfo {author} {\bibfnamefont {S.~N.}\ \bibnamefont {Filippov}}\ and\ \bibinfo {author} {\bibfnamefont {M.}~\bibnamefont {Ziman}},\ }\href {\doibase 10.1103/PhysRevA.88.032316} {\bibfield  {journal} {\bibinfo  {journal} {Phys. Rev. A}\ }\textbf {\bibinfo {volume} {88}},\ \bibinfo {pages} {032316} (\bibinfo {year} {2013})}\BibitemShut {NoStop}%
\bibitem [{\citenamefont {Masanes}\ \emph {et~al.}(2008)\citenamefont {Masanes}, \citenamefont {Liang},\ and\ \citenamefont {Doherty}}]{Masanes_PRL_2008}%
  \BibitemOpen
  \bibfield  {author} {\bibinfo {author} {\bibfnamefont {L.}~\bibnamefont {Masanes}}, \bibinfo {author} {\bibfnamefont {Y.-C.}\ \bibnamefont {Liang}}, \ and\ \bibinfo {author} {\bibfnamefont {A.~C.}\ \bibnamefont {Doherty}},\ }\href {\doibase 10.1103/PhysRevLett.100.090403} {\bibfield  {journal} {\bibinfo  {journal} {Phys. Rev. Lett.}\ }\textbf {\bibinfo {volume} {100}},\ \bibinfo {pages} {090403} (\bibinfo {year} {2008})}\BibitemShut {NoStop}%
\bibitem [{\citenamefont {Palazuelos}(2012)}]{Palazuelos_PRL_2012}%
  \BibitemOpen
  \bibfield  {author} {\bibinfo {author} {\bibfnamefont {C.}~\bibnamefont {Palazuelos}},\ }\href {\doibase 10.1103/PhysRevLett.109.190401} {\bibfield  {journal} {\bibinfo  {journal} {Phys. Rev. Lett.}\ }\textbf {\bibinfo {volume} {109}},\ \bibinfo {pages} {190401} (\bibinfo {year} {2012})}\BibitemShut {NoStop}%
\bibitem [{\citenamefont {Liang}\ \emph {et~al.}(2012)\citenamefont {Liang}, \citenamefont {Masanes},\ and\ \citenamefont {Rosset}}]{Liang_PRA_2012}%
  \BibitemOpen
  \bibfield  {author} {\bibinfo {author} {\bibfnamefont {Y.-C.}\ \bibnamefont {Liang}}, \bibinfo {author} {\bibfnamefont {L.}~\bibnamefont {Masanes}}, \ and\ \bibinfo {author} {\bibfnamefont {D.}~\bibnamefont {Rosset}},\ }\href {\doibase 10.1103/PhysRevA.86.052115} {\bibfield  {journal} {\bibinfo  {journal} {Phys. Rev. A}\ }\textbf {\bibinfo {volume} {86}},\ \bibinfo {pages} {052115} (\bibinfo {year} {2012})}\BibitemShut {NoStop}%
\bibitem [{\citenamefont {Quintino}\ \emph {et~al.}(2016)\citenamefont {Quintino}, \citenamefont {Brunner},\ and\ \citenamefont {Huber}}]{Quintino_PRA_2016}%
  \BibitemOpen
  \bibfield  {author} {\bibinfo {author} {\bibfnamefont {M.~T.}\ \bibnamefont {Quintino}}, \bibinfo {author} {\bibfnamefont {N.}~\bibnamefont {Brunner}}, \ and\ \bibinfo {author} {\bibfnamefont {M.}~\bibnamefont {Huber}},\ }\href {\doibase 10.1103/PhysRevA.94.062123} {\bibfield  {journal} {\bibinfo  {journal} {Phys. Rev. A}\ }\textbf {\bibinfo {volume} {94}},\ \bibinfo {pages} {062123} (\bibinfo {year} {2016})}\BibitemShut {NoStop}%
\bibitem [{\citenamefont {Hsieh}\ \emph {et~al.}(2016)\citenamefont {Hsieh}, \citenamefont {Liang},\ and\ \citenamefont {Lee}}]{Hsieh_PRA_2016}%
  \BibitemOpen
  \bibfield  {author} {\bibinfo {author} {\bibfnamefont {C.-Y.}\ \bibnamefont {Hsieh}}, \bibinfo {author} {\bibfnamefont {Y.-C.}\ \bibnamefont {Liang}}, \ and\ \bibinfo {author} {\bibfnamefont {R.-K.}\ \bibnamefont {Lee}},\ }\href {\doibase 10.1103/PhysRevA.94.062120} {\bibfield  {journal} {\bibinfo  {journal} {Phys. Rev. A}\ }\textbf {\bibinfo {volume} {94}},\ \bibinfo {pages} {062120} (\bibinfo {year} {2016})}\BibitemShut {NoStop}%
\bibitem [{\citenamefont {de~Vicente}(2014)}]{Vicente_JPA_2014}%
  \BibitemOpen
  \bibfield  {author} {\bibinfo {author} {\bibfnamefont {J.~I.}\ \bibnamefont {de~Vicente}},\ }\href {\doibase 10.1088/1751-8113/47/42/424017} {\bibfield  {journal} {\bibinfo  {journal} {Journal of Physics A: Mathematical and Theoretical}\ }\textbf {\bibinfo {volume} {47}},\ \bibinfo {pages} {424017} (\bibinfo {year} {2014})}\BibitemShut {NoStop}%
\bibitem [{\citenamefont {Brunner}\ \emph {et~al.}(2014)\citenamefont {Brunner}, \citenamefont {Cavalcanti}, \citenamefont {Pironio}, \citenamefont {Scarani},\ and\ \citenamefont {Wehner}}]{Brunner_RMP_2014}%
  \BibitemOpen
  \bibfield  {author} {\bibinfo {author} {\bibfnamefont {N.}~\bibnamefont {Brunner}}, \bibinfo {author} {\bibfnamefont {D.}~\bibnamefont {Cavalcanti}}, \bibinfo {author} {\bibfnamefont {S.}~\bibnamefont {Pironio}}, \bibinfo {author} {\bibfnamefont {V.}~\bibnamefont {Scarani}}, \ and\ \bibinfo {author} {\bibfnamefont {S.}~\bibnamefont {Wehner}},\ }\href {\doibase 10.1103/RevModPhys.86.419} {\bibfield  {journal} {\bibinfo  {journal} {Rev. Mod. Phys.}\ }\textbf {\bibinfo {volume} {86}},\ \bibinfo {pages} {419} (\bibinfo {year} {2014})}\BibitemShut {NoStop}%
\bibitem [{\citenamefont {Gallego}\ and\ \citenamefont {Aolita}(2015)}]{Gallego_PRX_2015}%
  \BibitemOpen
  \bibfield  {author} {\bibinfo {author} {\bibfnamefont {R.}~\bibnamefont {Gallego}}\ and\ \bibinfo {author} {\bibfnamefont {L.}~\bibnamefont {Aolita}},\ }\href {\doibase 10.1103/PhysRevX.5.041008} {\bibfield  {journal} {\bibinfo  {journal} {Phys. Rev. X}\ }\textbf {\bibinfo {volume} {5}},\ \bibinfo {pages} {041008} (\bibinfo {year} {2015})}\BibitemShut {NoStop}%
\bibitem [{\citenamefont {Uola}\ \emph {et~al.}(2020)\citenamefont {Uola}, \citenamefont {Costa}, \citenamefont {Nguyen},\ and\ \citenamefont {G\"uhne}}]{Uola_RMP_2020}%
  \BibitemOpen
  \bibfield  {author} {\bibinfo {author} {\bibfnamefont {R.}~\bibnamefont {Uola}}, \bibinfo {author} {\bibfnamefont {A.~C.~S.}\ \bibnamefont {Costa}}, \bibinfo {author} {\bibfnamefont {H.~C.}\ \bibnamefont {Nguyen}}, \ and\ \bibinfo {author} {\bibfnamefont {O.}~\bibnamefont {G\"uhne}},\ }\href {\doibase 10.1103/RevModPhys.92.015001} {\bibfield  {journal} {\bibinfo  {journal} {Rev. Mod. Phys.}\ }\textbf {\bibinfo {volume} {92}},\ \bibinfo {pages} {015001} (\bibinfo {year} {2020})}\BibitemShut {NoStop}%
\bibitem [{\citenamefont {Hsieh}(2020)}]{Hsieh_Quantum_2020}%
  \BibitemOpen
  \bibfield  {author} {\bibinfo {author} {\bibfnamefont {C.-Y.}\ \bibnamefont {Hsieh}},\ }\href {\doibase 10.22331/q-2020-03-19-244} {\bibfield  {journal} {\bibinfo  {journal} {Quantum}\ }\textbf {\bibinfo {volume} {4}},\ \bibinfo {pages} {244} (\bibinfo {year} {2020})}\BibitemShut {NoStop}%
\bibitem [{\citenamefont {Vedral}\ \emph {et~al.}(1997)\citenamefont {Vedral}, \citenamefont {Plenio}, \citenamefont {Rippin},\ and\ \citenamefont {Knight}}]{Vidal_PRL_1997}%
  \BibitemOpen
  \bibfield  {author} {\bibinfo {author} {\bibfnamefont {V.}~\bibnamefont {Vedral}}, \bibinfo {author} {\bibfnamefont {M.~B.}\ \bibnamefont {Plenio}}, \bibinfo {author} {\bibfnamefont {M.~A.}\ \bibnamefont {Rippin}}, \ and\ \bibinfo {author} {\bibfnamefont {P.~L.}\ \bibnamefont {Knight}},\ }\href {\doibase 10.1103/PhysRevLett.78.2275} {\bibfield  {journal} {\bibinfo  {journal} {Phys. Rev. Lett.}\ }\textbf {\bibinfo {volume} {78}},\ \bibinfo {pages} {2275} (\bibinfo {year} {1997})}\BibitemShut {NoStop}%
\bibitem [{\citenamefont {Stratton}\ \emph {et~al.}(2024)\citenamefont {Stratton}, \citenamefont {Hsieh},\ and\ \citenamefont {Skrzypczyk}}]{Stratton_PRL_2024}%
  \BibitemOpen
  \bibfield  {author} {\bibinfo {author} {\bibfnamefont {B.}~\bibnamefont {Stratton}}, \bibinfo {author} {\bibfnamefont {C.-Y.}\ \bibnamefont {Hsieh}}, \ and\ \bibinfo {author} {\bibfnamefont {P.}~\bibnamefont {Skrzypczyk}},\ }\href {\doibase 10.1103/PhysRevLett.132.110202} {\bibfield  {journal} {\bibinfo  {journal} {Phys. Rev. Lett.}\ }\textbf {\bibinfo {volume} {132}},\ \bibinfo {pages} {110202} (\bibinfo {year} {2024})}\BibitemShut {NoStop}%
\bibitem [{\citenamefont {Chitambar}\ \emph {et~al.}(2014)\citenamefont {Chitambar}, \citenamefont {Leung}, \citenamefont {Mančinska}, \citenamefont {Ozols},\ and\ \citenamefont {Winter}}]{Chitambar_CMP_2014}%
  \BibitemOpen
  \bibfield  {author} {\bibinfo {author} {\bibfnamefont {E.}~\bibnamefont {Chitambar}}, \bibinfo {author} {\bibfnamefont {D.}~\bibnamefont {Leung}}, \bibinfo {author} {\bibfnamefont {L.}~\bibnamefont {Mančinska}}, \bibinfo {author} {\bibfnamefont {M.}~\bibnamefont {Ozols}}, \ and\ \bibinfo {author} {\bibfnamefont {A.}~\bibnamefont {Winter}},\ }\href {\doibase 10.1007/s00220-014-1953-9} {\bibfield  {journal} {\bibinfo  {journal} {Communications in Mathematical Physics}\ }\textbf {\bibinfo {volume} {328}},\ \bibinfo {pages} {303} (\bibinfo {year} {2014})}\BibitemShut {NoStop}%
\bibitem [{\citenamefont {Moravčíková}\ and\ \citenamefont {Ziman}(2010{\natexlab{b}})}]{Moravcikova_JPA_2010}%
  \BibitemOpen
  \bibfield  {author} {\bibinfo {author} {\bibfnamefont {L.}~\bibnamefont {Moravčíková}}\ and\ \bibinfo {author} {\bibfnamefont {M.}~\bibnamefont {Ziman}},\ }\href {\doibase 10.1088/1751-8113/43/27/275306} {\bibfield  {journal} {\bibinfo  {journal} {Journal of Physics A: Mathematical and Theoretical}\ }\textbf {\bibinfo {volume} {43}},\ \bibinfo {pages} {275306} (\bibinfo {year} {2010}{\natexlab{b}})}\BibitemShut {NoStop}%
\bibitem [{\citenamefont {Gour}\ and\ \citenamefont {Scandolo}(2020)}]{Gour_arXiv_2020}%
  \BibitemOpen
  \bibfield  {author} {\bibinfo {author} {\bibfnamefont {G.}~\bibnamefont {Gour}}\ and\ \bibinfo {author} {\bibfnamefont {C.~M.}\ \bibnamefont {Scandolo}},\ }\href {https://arxiv.org/abs/2101.01552} {\bibfield  {journal} {\bibinfo  {journal} {arXiv: 2101.01552}\ } (\bibinfo {year} {2020})}\BibitemShut {NoStop}%
\bibitem [{\citenamefont {Monroe}\ \emph {et~al.}(2014)\citenamefont {Monroe}, \citenamefont {Raussendorf}, \citenamefont {Ruthven}, \citenamefont {Brown}, \citenamefont {Maunz}, \citenamefont {Duan},\ and\ \citenamefont {Kim}}]{monroe2014large}%
  \BibitemOpen
  \bibfield  {author} {\bibinfo {author} {\bibfnamefont {C.}~\bibnamefont {Monroe}}, \bibinfo {author} {\bibfnamefont {R.}~\bibnamefont {Raussendorf}}, \bibinfo {author} {\bibfnamefont {A.}~\bibnamefont {Ruthven}}, \bibinfo {author} {\bibfnamefont {K.~R.}\ \bibnamefont {Brown}}, \bibinfo {author} {\bibfnamefont {P.}~\bibnamefont {Maunz}}, \bibinfo {author} {\bibfnamefont {L.-M.}\ \bibnamefont {Duan}}, \ and\ \bibinfo {author} {\bibfnamefont {J.}~\bibnamefont {Kim}},\ }\href {\doibase 10.1103/PhysRevA.89.022317} {\bibfield  {journal} {\bibinfo  {journal} {Physical Review A}\ }\textbf {\bibinfo {volume} {89}},\ \bibinfo {pages} {022317} (\bibinfo {year} {2014})}\BibitemShut {NoStop}%
\bibitem [{\citenamefont {Van~Meter}\ and\ \citenamefont {Devitt}(2016)}]{van2016path}%
  \BibitemOpen
  \bibfield  {author} {\bibinfo {author} {\bibfnamefont {R.}~\bibnamefont {Van~Meter}}\ and\ \bibinfo {author} {\bibfnamefont {S.~J.}\ \bibnamefont {Devitt}},\ }\href {\doibase 10.1109/MC.2016.291} {\bibfield  {journal} {\bibinfo  {journal} {Computer}\ }\textbf {\bibinfo {volume} {49}},\ \bibinfo {pages} {31} (\bibinfo {year} {2016})}\BibitemShut {NoStop}%
\bibitem [{\citenamefont {Fitzsimons}(2017)}]{fitzsimons2017private}%
  \BibitemOpen
  \bibfield  {author} {\bibinfo {author} {\bibfnamefont {J.~F.}\ \bibnamefont {Fitzsimons}},\ }\href {\doibase https://doi.org/10.1038/s41534-017-0025-3} {\bibfield  {journal} {\bibinfo  {journal} {npj Quantum Information}\ }\textbf {\bibinfo {volume} {3}},\ \bibinfo {pages} {23} (\bibinfo {year} {2017})}\BibitemShut {NoStop}%
\bibitem [{\citenamefont {Takeuchi}\ \emph {et~al.}(2016)\citenamefont {Takeuchi}, \citenamefont {Fujii}, \citenamefont {Ikuta}, \citenamefont {Yamamoto},\ and\ \citenamefont {Imoto}}]{takeuchi2016blind}%
  \BibitemOpen
  \bibfield  {author} {\bibinfo {author} {\bibfnamefont {Y.}~\bibnamefont {Takeuchi}}, \bibinfo {author} {\bibfnamefont {K.}~\bibnamefont {Fujii}}, \bibinfo {author} {\bibfnamefont {R.}~\bibnamefont {Ikuta}}, \bibinfo {author} {\bibfnamefont {T.}~\bibnamefont {Yamamoto}}, \ and\ \bibinfo {author} {\bibfnamefont {N.}~\bibnamefont {Imoto}},\ }\href {\doibase 10.1103/PhysRevA.93.052307} {\bibfield  {journal} {\bibinfo  {journal} {Physical Review A}\ }\textbf {\bibinfo {volume} {93}},\ \bibinfo {pages} {052307} (\bibinfo {year} {2016})}\BibitemShut {NoStop}%
\bibitem [{\citenamefont {Sheng}\ and\ \citenamefont {Zhou}(2018)}]{sheng2018blind}%
  \BibitemOpen
  \bibfield  {author} {\bibinfo {author} {\bibfnamefont {Y.-B.}\ \bibnamefont {Sheng}}\ and\ \bibinfo {author} {\bibfnamefont {L.}~\bibnamefont {Zhou}},\ }\href {\doibase 10.1103/PhysRevA.98.052343} {\bibfield  {journal} {\bibinfo  {journal} {Physical Review A}\ }\textbf {\bibinfo {volume} {98}},\ \bibinfo {pages} {052343} (\bibinfo {year} {2018})}\BibitemShut {NoStop}%
\end{thebibliography}%
\end{document}